%% file: ghz_v2.tex
\documentclass[aps,prl,groupedaddress,twocolumn]{revtex4-1}

\usepackage{graphicx}  
\usepackage{amssymb}   
\usepackage{amsmath}
\usepackage{physics}
\usepackage[usenames,dvipsnames]{color}
\usepackage[colorlinks=true, citecolor=Red,urlcolor=blue, linkcolor=BrickRed]{hyperref}

\input{Qcircuit.tex}

\renewcommand{\emph}{\textit}

\begin{document}

\title{Verifying Multipartite Entangled GHZ States via Multiple Quantum Coherences}

\author{Ken X. Wei}
\email{xkwei@ibm.com}
\author{Isaac Lauer}
\author{Srikanth Srinivasan}
\author{Neereja Sundaresan}
\author{Douglas T. McClure}
\author{David Toyli}
\author{David C. McKay}
\author{Jay M. Gambetta}
\author{Sarah Sheldon}
\affiliation{IBM T.J. Watson Research Center, Yorktown Heights, NY 10598, USA}

\date{\today}

\begin{abstract}
The ability to generate and verify multipartite entanglement is an important benchmark for near-term quantum devices. We develop a scalable entanglement metric based on multiple quantum coherences, and demonstrate experimentally on a 20-qubit superconducting device. We report a state fidelity of $0.5165\pm0.0036$ for an 18-qubit GHZ state, indicating multipartite entanglement across all 18 qubits. Our entanglement metric is robust to noise and only requires measuring the population in the ground state; it can be readily applied to other quantum devices to verify multipartite entanglement.
\end{abstract}

\maketitle

Universal quantum computers promise to solve many problems that are intractable classically~\cite{shor96, Bravyi308}, but achieving fault tolerance will require a number of resources that are unavailable today. Until we can implement error correction, quantum systems will be beset with a certain amount of noise. Understanding how to best benchmark these near-term quantum devices is an active question~\cite{Preskill2018}. Traditionally, the field has relied on local metrics such as one- and two-qubit gate fidelities since these are experimentally feasible even with full tomographic methods~\cite{knill2008, Magesan2012, Magesan2012a, sheldonIRB}. However, it has become increasingly clear that such local metrics do not capture the full intricacies of a multi-qubit device. Therefore, a number of multi-qubit metrics such as direct fidelity estimation~\cite{FL,SLCP}, three qubit simultaneous randomized benchmarking (RB)~\cite{McKay17}, direct RB~\cite{proctor18}, and quantum volume~\cite{Cross2018} have been proposed and measured. Another powerful multi-qubit metric is that of entanglement, specifically, measuring the largest possible multipartite entangled state on a device~\cite{Wang2018, mooney19x}. Not only is the ability to generate entanglement indicative of high fidelity gate operations and qubit coherence, entangled states are the cornerstone of quantum speedups and they can be direct resources for quantum computing~\cite{shor97,raussendorf01l}.  Multipartite entanglement in Greenberger-Horne-Zeilinger (GHZ) states have been demonstrated with 10 superconducting circuits~\cite{Song17}, 14 trapped ions~\cite{Monz2011}, and 18 photons~\cite{wang18l}. Recently, multipartite entanglement in a 12-qubit linear graph state was verified in a superconducting qubit architecture~\cite{gong19l}. 

\begin{figure}[t]
\includegraphics[width = 0.75\columnwidth]{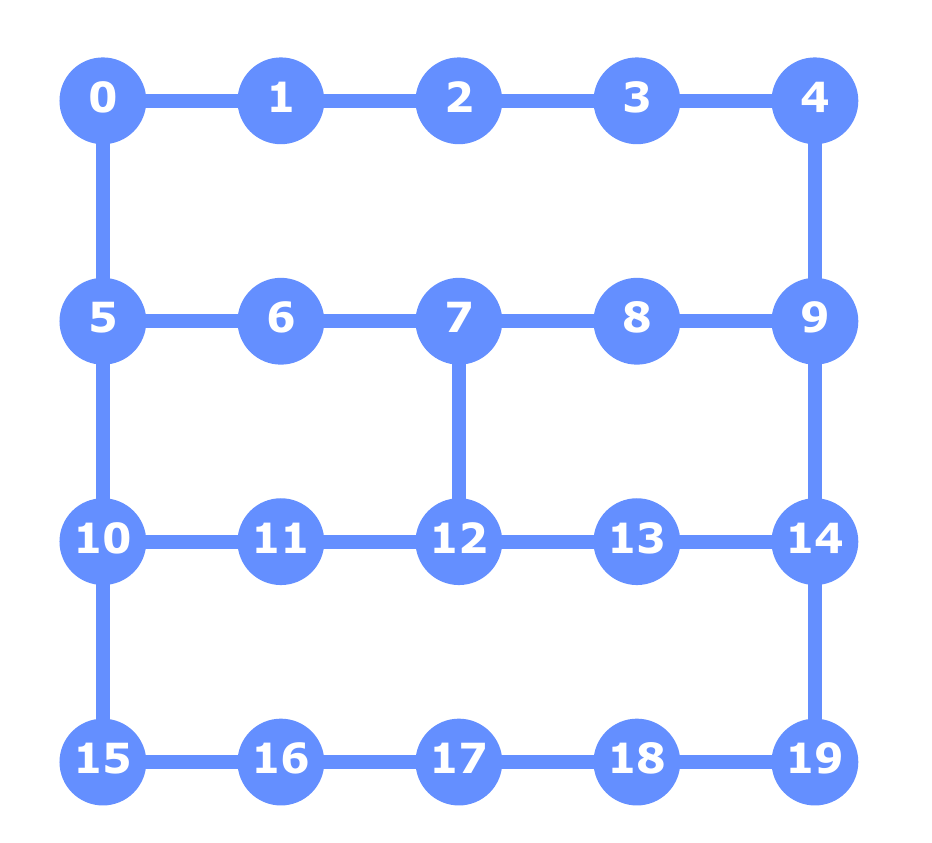}
\includegraphics[width = \columnwidth]{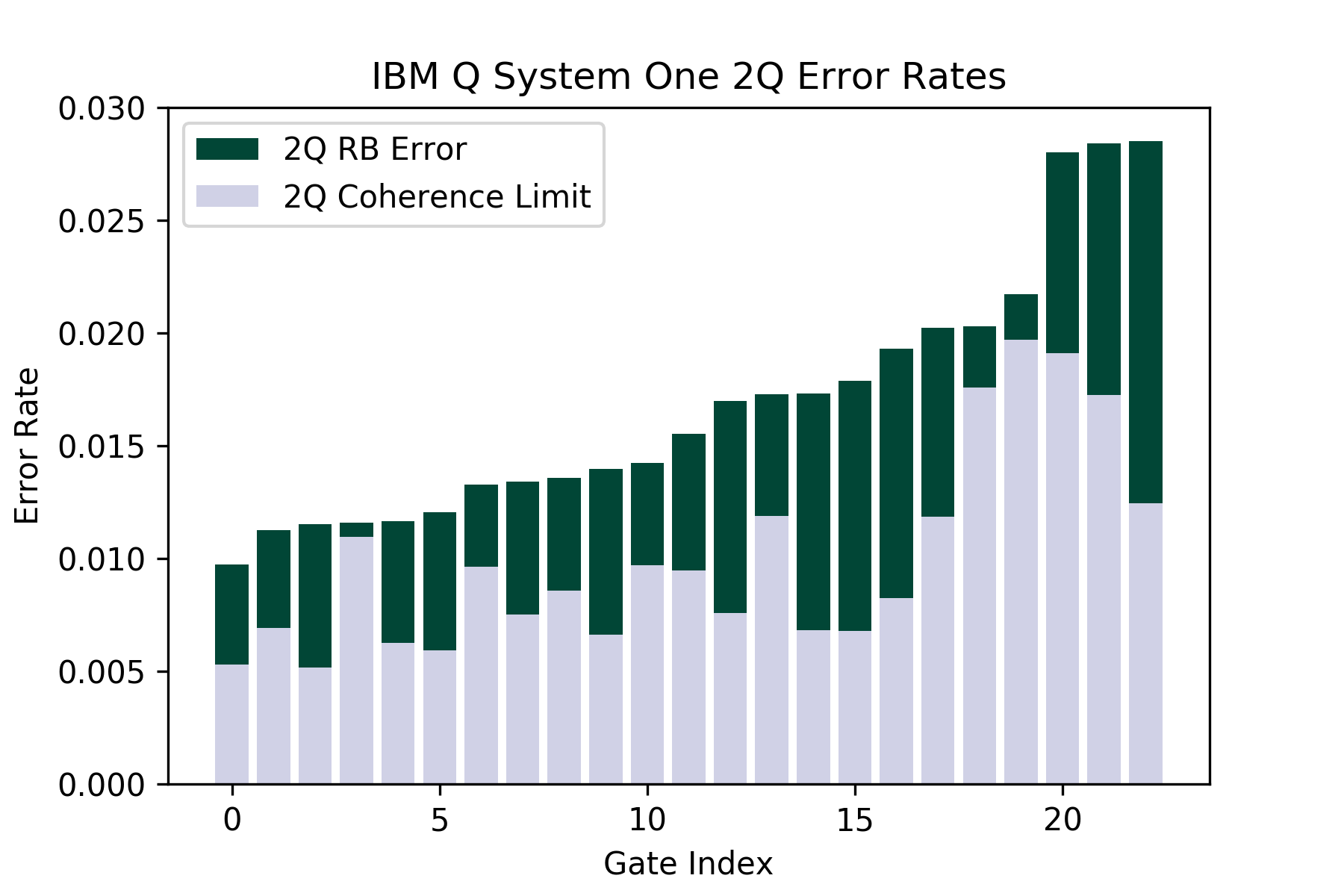}
\caption{\textbf{System One device layout and 2Q errors} Top: 20 qubit device layout and connectivity on IBM Q System One. Bottom: Comparison between 2Q error and 2Q coherence limit for all two-qubit gates on the device. The coherence properties of the device can be found in~\cite{qvibm}.}
\label{fig:ibmq}
\end{figure}

Here we generate and verify an 18-qubit entangled GHZ state on a 20-qubit superconducting device. Our entanglement metric is inspired by quantum sensing~\cite{degen17rmp} and can be used to directly bound the state fidelity. The device is comprised of 20 fixed frequency transmon qubits, and implements two-qubit gates based on cross-resonance driving~\cite{chow_CR_2011, rigetti, sheldonCR}. The device layout and the two qubit errors are shown in FIG.~\ref{fig:ibmq}.

\begin{figure*}[!bht]
\minipage{0.45\textwidth}
\centering
\scriptsize
\input{circ1}

\input{circ2}

\input{circ3}

\endminipage
\minipage{0.55\textwidth}
\centering
\scriptsize
\input{circ4}
\endminipage
\caption{\textbf{Quantum circuits} {\textbf{A}}: quantum sensing circuit. An ideal GHZ state is generated and used to sense an external magnetic field. After sensing, the GHZ state is disentangled and information about the magnetic field is encoded as a phase on the first qubit. {\textbf{B}}: MQC quantum circuit. Instead of sensing, we apply a collective rotation given by the unitary operator $U_\phi = e^{-i\phi/2\sum_j \sigma_z^j}$. We can implement this rotation instantaneously in our device by phase-shifting all subsequent pulses~\cite{McKay2016b}. In the readout step, all qubits are measured to obtain the probability of system returning to initial state. {\textbf{C}}: refocused MQC quantum circuit. Similar to MQC except for the addition of a collective $\pi$-pulse on all qubits before $U_\phi$. The $\pi$-pulse is used to reduce noise without affecting the GHZ state. {\textbf{D}}: experimental circuit for the 18-qubit MQC experiment on IBM Q System One.}
\label{fig:circ}
\end{figure*}
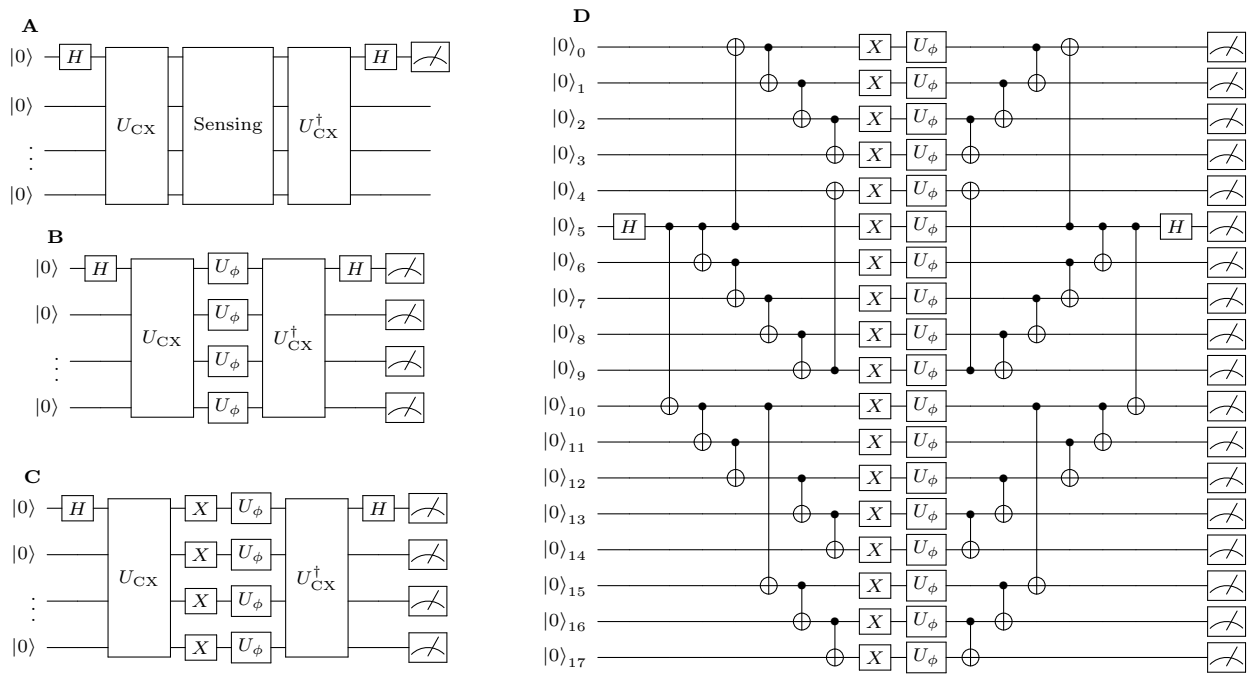

We verify the generation of GHZ states by measuring multiple quantum coherences (MQC)~\cite{baum85jcp}, a tool traditionally used in solid state NMR and more recently in trapped ions to study many-body correlations and quantum information scrambling~\cite{wei18prl, garttner17nphy}. The experimental method to measure MQC has a strong overlap with quantum sensing and entanglement assisted metrology~\cite{cappellaro05prl}. In the prototypical quantum sensing circuit shown in FIG.~\ref{fig:circ}{\textbf{A}}, a GHZ state is used to sense static magnetic fields with Heisenberg-limited sensitivity; it works by taking advantage of an ideal GHZ state's amplified sensitivity to phase rotations of each of the individual qubits in the entangled state~\cite{bollinger96a, Leibfried1476, Giovannetti1330, degen17rmp}. If each qubit has a phase rotation of $\phi$, then the $N$-qubit GHZ state rotates collectively by $N\phi$. By observing how sensitive a nonideal GHZ state responds to rotations, we can deduce how much entanglement is present in the state. 
The quantum circuit for measuring MQC is illustrated in FIG.~\ref{fig:circ}{\textbf{B}}, and it can be described in four steps:
\begin{enumerate}
\item Starting from the $N$-qubit ground state: $\ket{\text{GS}}=\ket{00..000}$, apply a Hadamard gate on qubit 0 followed by a sequence of CX gates. Ideally this brings the system into the GHZ state: $\ket{\text{GHZ}}=\frac{1}{\sqrt{2}}(\ket{000..00}+\ket{111..11})$
\item Apply a collective rotation given by the unitary $U_\phi$ on all qubits. This amounts to adding a phase $N \phi$ to the GHZ state: $\frac{1}{\sqrt{2}}(\ket{000..00}+e^{-i N \phi} \ket{111..11})$
\item Disentangle the GHZ state by performing the CX gate sequence in reverse order. The amplified phase is mapped onto qubit 0: $\frac{1}{\sqrt{2}}(\ket{0}+e^{-i N \phi} \ket{1})\otimes\ket{00..00}$
\item Read out the amplified phase by measuring the probability of the system returning to its initial state: $\ket{\text{GS}}$
\end{enumerate}
The measured signal of this protocol is given by
\begin{align}
S_\phi=|\bra{000...00}U^\dagger_\text{GHZ} U_\phi U_\text{GHZ}\ket{000...00}|^2 = \text{Tr}(\rho_\phi \rho)
\label{eq:overlap}
\end{align}
where $\rho=U_\text{GHZ}\ketbra{\text{GS}}{\text{GS}}U_\text{GHZ}^\dagger$, $U_\text{GHZ}=U_\text{CX}H_{0}$, and $\rho_\phi = U_\phi \rho U_\phi^\dagger$.
If our controls are perfect and there is no decoherence, Eq.~(\ref{eq:overlap}) reduces to 
\begin{align*}
S^\text{ideal}_\phi=\frac{1}{2}(1+\cos(N \phi))
\end{align*}
which can also be obtained by measuring the state $\ket{0}$ on qubit 0 in the final step. Since $S_\phi$ comes from the overlap between a rotated and unrotated density matrix, in the final step of the protocol the probability of all qubits being in the zero state
must be measured. The constant term in $S^\text{ideal}$ comes from the diagonal elements of the GHZ density matrix, whereas the oscillating term comes from the off-diagonal corner elements. Any difference between $S_\phi$ and $S_\phi^\text{ideal}$ is an indication that our GHZ state is imperfect. To quantify the state fidelity we focus on the MQC amplitudes, defined as the discrete Fourier transform of $S_\phi$: 
\begin{align*}
I_q = \mathcal{N}^{-1} |\sum_\phi e^{iq\phi}S_\phi|
\end{align*}
where $\mathcal{N}$ is a normalization factor. 
The $N$-qubit GHZ state fidelity defined as $F=\mel{\text{GHZ}}{\rho}{\text{GHZ}}$ can be bounded by
\begin{align}\label{eq:fbound}
2\sqrt{I_N} \leq F \leq \sqrt{I_0/2} + \sqrt{I_N}
\end{align}
For a perfect GHZ state we have $I_0 = 2I_N = 1/2$, and all other $I_q$ being zero. We can also directly obtain the state fidelity as $F=\frac{1}{2}(P_{000..00}+P_{111..11})+\sqrt{I_N}$, where $P_{000..00}$ and $P_{111..11}$ are the populations of $\ket{000..00}$ and $\ket{111..11}$ in the density matrix. A discussion on MQC amplitudes and proof of Eq.~(\ref{eq:fbound})
are given in SM. For a $N$-qubit state to have multipartite GHZ entanglement, it needs to have a minimal fidelity of 0.5~\cite{guhne09pr, Guhne_njp}. 
%

%
%
%
%

\begin{figure*}[t]
\begin{minipage}{0.48\textwidth}
\centering
\includegraphics[width = \columnwidth]{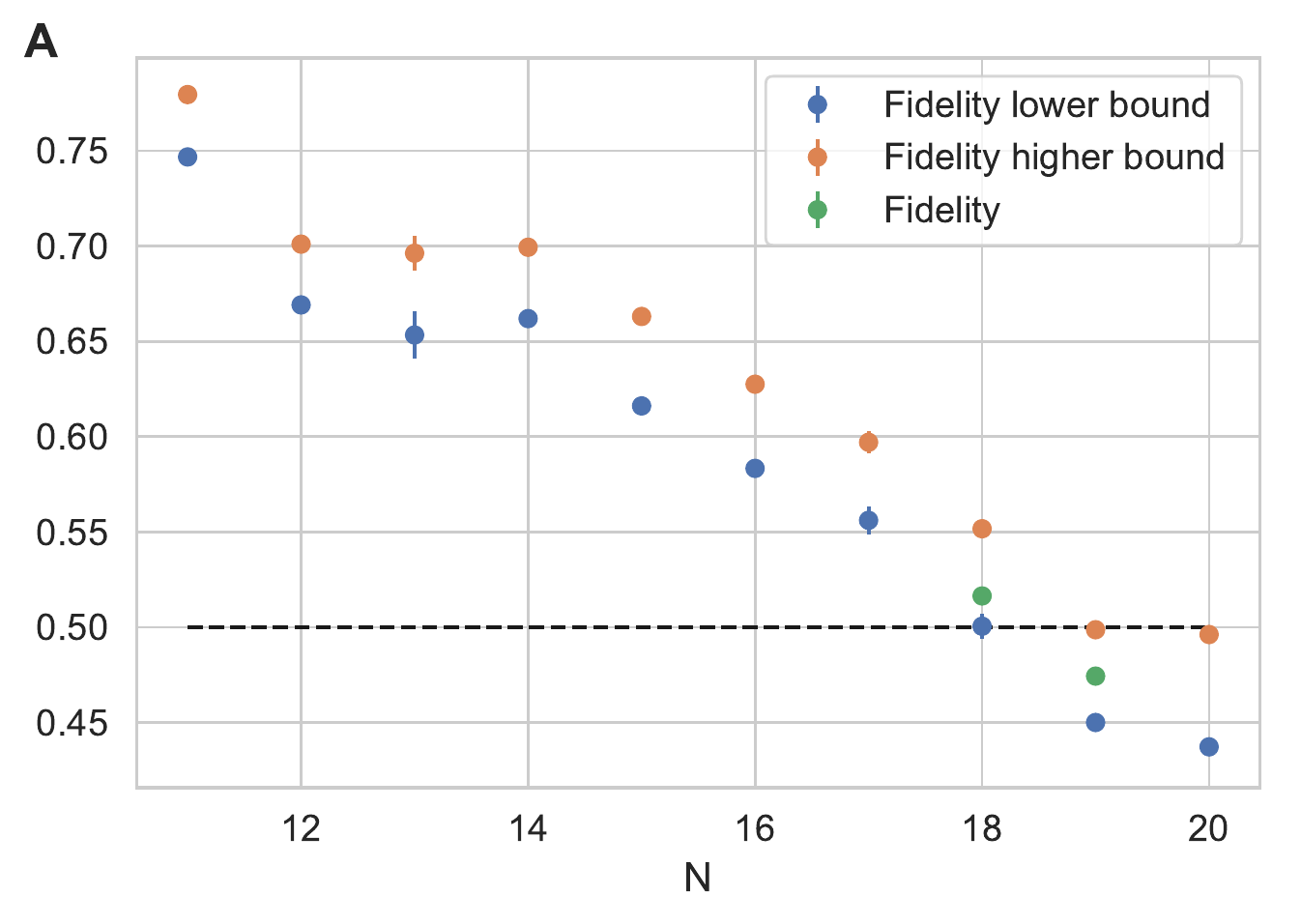}
\end{minipage}
\begin{minipage}{0.48\textwidth}
\centering
\includegraphics[width = \columnwidth]{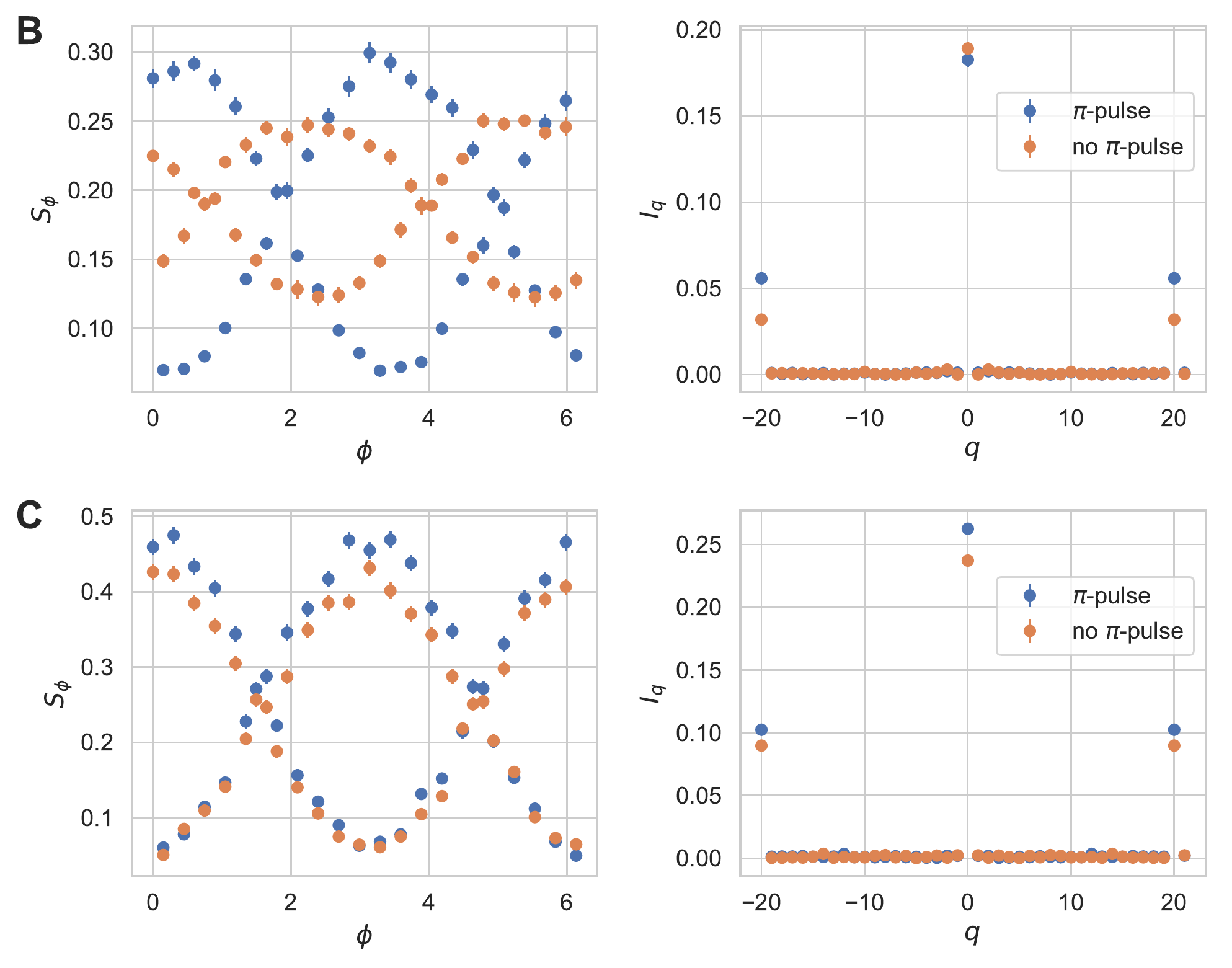}
\end{minipage}
\caption{\textbf{Fidelity bounds and 20q data} ${\bf A}$: experimentally extracted fidelity bounds according to Eq.~(\ref{eq:fbound}) for $N=11$ to $N=20$ with readout correction. Error bars corresponds to linearly propagated uncertainties in the corresponding $S_\phi$. For $N=18$ and $N=19$ the fidelities are also presented. ${\bf B}$: experimentally measured 20-qubit $S_\phi$ and MQC amplitudes with and without the refocusing $\pi$-pulse. ${\bf C}$: simulated results using basic noise model described in~\cite{Qiskit19}}.
\label{fig:fbound}
\end{figure*}

While this method departs from parity oscillation measurements commonly used in trapped ions to verify GHZ entanglement~\cite{sackett00n, leibfried05n, Monz2011}, MQC offers two main benefits: robustness to noise and scalability in readout correction. Parity oscillations measure the coherence between the $\ket{000..00}$ and $\ket{111..11}$ states. It works by looking at the oscillations in the expectation value $\expval{ZZZ ... ZZ}{\text{GHZ}_\phi}$ 
as a function of $\phi$, where $\text{GHZ}_\phi = \otimes^N_j e^{\frac{i\pi}{4} (\cos\phi \sigma_x^j + \sin\phi \sigma_y^j )}\ket{\text{GHZ}}$~\cite{Monz2011}. The amplitude of the parity oscillations gives coherence $C=|\rho_{000..00,111..11}|+|\rho_{111..11,000..00}|$, which is related to the fidelity via $F=\frac{1}{2}(P_{000..00}+P_{111..11}+C)$~\cite{Monz2011}. The coherence is related to MQC amplitudes via $C=2\sqrt{I_N}$. At first our entanglement metric appears disadvantageous compared to parity oscillations since it takes twice the circuit length. However our experiments can be made robust against noise. Just as a Hahn echo refocuses low frequency noise and reduces dephasing~\cite{hahn50}, adding a $\pi$-pulse after making the GHZ state can dramatically improve the measured fidelity. The quantum circuit for refocused MQC is illustrated in FIG.~\ref{fig:circ}{\textbf{C}}. 
We find experimentally the 20-qubit state fidelity to increase by nearly $11\%$ by adding the refocusing $\pi$-pulse, as shown in FIG.~\ref{fig:fbound}{\textbf{B}}. In addition, our entanglement metric only requires measurement of the initial state.

In addition to accommodating dynamical decoupling techniques, the MQC method is also less sensitive to readout errors.  
We point out that parity oscillations in GHZ states have been measured previously on the IBM Q 16-qubit device~\cite{cruz18x} with average readout error of $7\%$, but multipartite entanglement cannot be established beyond five qubits. Aside from control imperfections and decoherence, readout errors limit our ability to measure the entangled state, even though the state itself can be highly entangled. Since readout errors are independent from entanglement we can calibrate them out of the measurement. To mitigate measurement errors, we experimentally construct a $2^N$ by $2^N$ calibration matrix, $A$, where each row vector corresponds to the measured outcome probabilities of a prepared basis state. 
In the nominal case of no readout error, $A$ is an identity matrix. With readout error, we can correct the measured counts $v_\text{mea}$ by minimizing
\begin{align}
|A v_\text{cal}- v_\text{mea}|^2
\label{eq:amat}
\end{align}
under the constraint $\sum_j v_{\text{cal},j} = 1$ and $v_{\text{cal},j} \geq 0$. Here $v_\text{cal}$ is the calibrated counts of $v_\text{mea}$. Eq.~(\ref{eq:amat}) can be recasted into a convex optimization problem
and solved by quadratic programming  using packages such as CVXOPT~\cite{cvxopt}. The overhead for measuring $A$ and minimizing Eq.~(\ref{eq:amat}) increases exponentially with $N$. We modify this calibration procedure to have a scalable way to perform readout correction using two key features of the MQC method. For one, we only need to measure the probability that the sate is in $\ket{000..00}$. Two, we expect that the MQC output for imperfections in the GHZ state to result in low excitation states.
This is not entirely unexpected since for an ideal GHZ state, the MQC output has only two distinct states: $\ket{000..00}$ and $\ket{100..00}$. With an imperfect GHZ state and readout errors we expect the output counts to spread out but stay within the low excitation manifolds. Combining these two features we can significantly reduce the overhead for readout calibration by truncating $A$ to only correct for states with significant weights. For example, we can reasonably conclude that the final measurement will not include states such as $\ket{111..11}$. We justify this truncation in SM, where we present the corrected data as a function of the number of states kept in the calibration. A rapid convergence is observed with just 256 states. While keeping such a low number of states may not be sufficient to correct for the entire output vector, it is sufficient to accurately correct for the $\ket{000..00}$ state. In addition, the largest 256 states all have similar excitation distributions centered around three excitations independent of $N$, as shown in SM. This suggests the readout calibration based on the truncated $A$, call it $A_t$, is scalable for measuring MQC amplitudes in GHZ states. The $A_t$ matrix cannot correct readout errors from parity oscillation experiments, since the output counts will be distributed across all eigenstates. An alternative to scalable readout calibration is to approximate the full $A$ matrix as a tensor product of $A$ matrices of each qubit~\cite{Song17,gong19l}, this approach is valid when there is little to no readout cross-talk between qubits. 

\begin{figure*}[!tbh]
\includegraphics[width=0.98\textwidth]{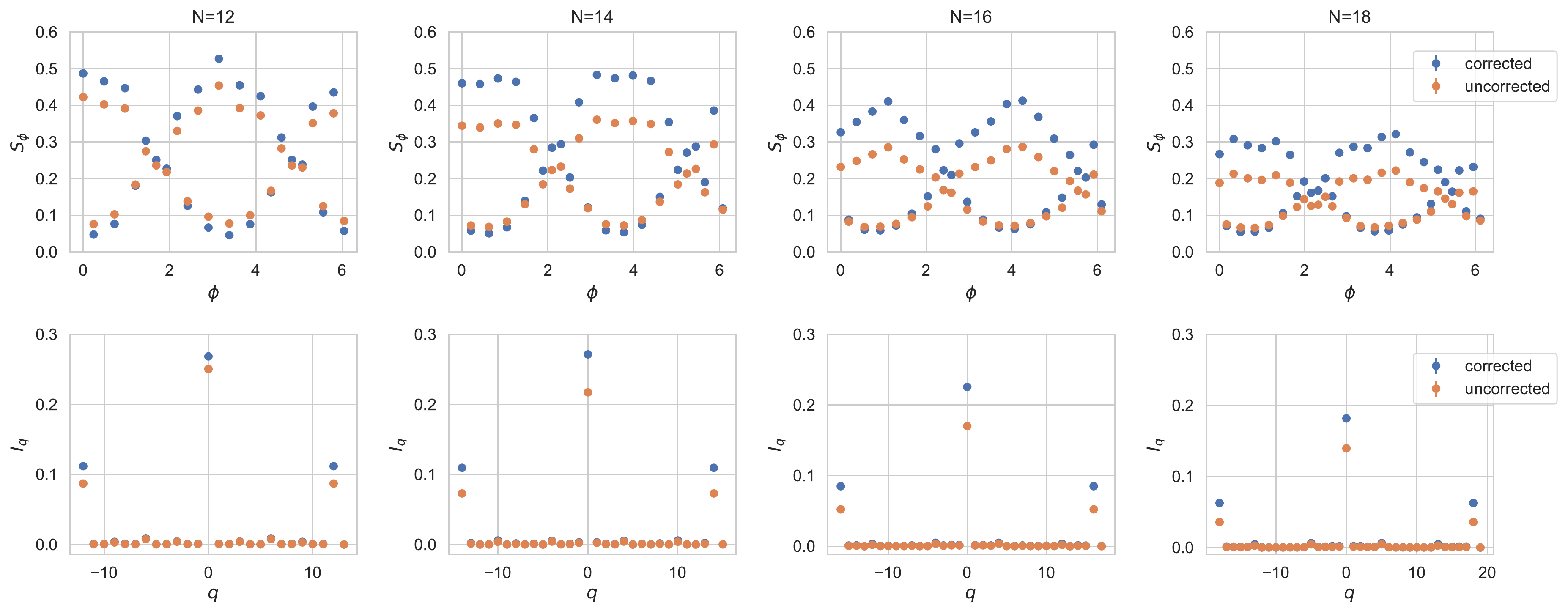}
\caption{\textbf{Experimentally measured $S_\phi$ and extracted MQC amplitudes} Top row: experimentally measured $S_\phi$ for $N=12,14,16,18$. Bottom row: corresponding MQC amplitudes extracted by discrete Fourier transforming $S_\phi$. The errorbar corresponds to one standard error linearly propagated from uncertainties in $S_\phi$.}
\label{fig:data}
\end{figure*}

We have experimentally generated GHZ states for $N=11$ to $N=20$, and measured $S_\phi$ in Eq.~(\ref{eq:overlap}) and extracted the corresponding MQC amplitudes. The data for 12, 14, 16, and 18-qubit GHZ states are shown in FIG.~\ref{fig:data}, and the circuit used in the 18-qubit experiment is shown in FIG.~\ref{fig:circ}{\textbf{D}}. For each $N$, we measure $S_\phi$ for $\phi=\frac{\pi j}{N+1}$, where $j=0,1,2,\cdots,2N+1$ so the highest frequency detectable is $N+1$. The result of each experiment is averaged over 16384 shots, and the errorbar corresponds to one standard error obtained from eight experiments. There is considerable difference between the results with and without readout correction. 
 Here $A_t$ is constructed using 256 basis states, and each basis state measurement is averaged over 4096 shots. 

\begin{table}[b]
\caption{\label{tab:ghzcomp}\textbf{Fidelity Comparison for small GHZ states}}
\begin{ruledtabular}
\begin{tabular}{lcccc}
Method & 2q GHZ & 3q GHZ & 4q GHZ & 5q GHZ\\
\colrule
QST & 0.96 & 0.93 & 0.87 & 0.85\\
MQC & 0.98 & 0.94 & 0.87 & 0.86\\
\end{tabular}
\end{ruledtabular}
\end{table}

From the experimentally extracted MQC amplitudes shown in the bottom row of FIG.~\ref{fig:data}, we see one peak located at $q=0$ and two peaks at $q=\pm N$, characteristic of $N$-qubit GHZ states. Peak amplitudes become lower with increasing $N$, indicating larger $N$-qubit GHZ states have lower fidelities. Using Eq.~(\ref{eq:fbound}) we extract the upper and lower bounds on state fidelities with readout calibrations as a function of $N$, as shown in FIG.~\ref{fig:fbound}{\textbf{A}}. For $N=11$ to $N=17$ the fidelity lower bound is clearly higher than the 0.5 threshold for multipartite entanglement. For $N=18$ the lower bound is $0.5006\pm0.0067$, in this case we measure $P_{000..00}$ and $P_{111..11}$ for the GHZ state in addition to MQC amplitudes to obtain the state fidelity of $F=0.5165\pm0.0036$, confirming that the 18-qubit GHZ state is multipartite entangled. We have not been able to establish multipartite entanglement with 19 and 20-qubit GHZ states. Without applying the aforementioned readout calibration, the highest number of multipartite entangled qubits we can measure is $N=14$ with a fidelity lower bound of $0.5406\pm0.0037$. We compare fidelities extracted from our method with that from quantum state tomography (QST)~\cite{Smolin12} for small GHZ states, the results are summarized in Table~\ref{tab:ghzcomp}. 
While the MQC method appears to give slightly higher state fidelity, we expect these results are within the errors of the tomography experiments. Each experiment is averaged over 16384 shots and readout corrected. Our method to experimentally quantify multipartite entanglement for GHZ states can be applied to other states that are locally equivalent to GHZ states, such as star graph and complete graph states~\cite{hein04a}. The only difference is in the rotation step. For star graph states, instead of applying $U_\phi$ on all qubits, apply $U_\phi$ on the central qubit and $H U_\phi H$ on the rest. For complete graph states, apply $e^{-\frac{i \pi}{4} \sigma_x} U_\phi e^{\frac{i \pi}{4} \sigma_x}$ on all qubits.

There are several experimental limitations to entangling large GHZ states in our device. First, the circuit depth required to generate a $N$-qubit GHZ state scales as $\mathcal{O}(N)$; in contrast with the linear graph state, where only two steps are needed independent of $N$~\cite{gong19l, mooney19x}. This makes GHZ states particularly fragile to decoherence. We choose an entangling path on our device that takes the least amount of time to complete; the physical qubits involved for each $N$ are listed in SM. Second, there are instances where CX gates, implemented by cross-resonance driving, are run on adjacent qubit pairs. Simultaneous adjacent CX gates will have lower fidelity than individual CX gates due to always-on $ZZ$ interactions and cross-driving effects between neighboring qubits. Third, since the CX gates are applied sequentially, there will inevitably be free evolution on idle qubits leading to unitary errors. In addition, due to pulse alignment restrictions in the software, the entangling and the disentangling operations take different times to complete, making the $\pi$-pulse not as effective as it can be. This might explain why the measured $S_\phi$ appears phase-shifted. 

We use a basic noise model built from device parameters to simulate the 20-qubit MQC experiments using Qiskit~\cite{Qiskit19}. The simulation models one- and two-qubit gate errors as a depolarizing error plus a thermal relaxation error such that the total error equals the error measured experimentally from randomized benchmarking; 
the details of the basic noise model can be found in Qiskit tutorial~\cite{Qiskit19}. To compare with experiments, we turn off readout error in the simulation and average over 2048 shots. The simulation shows higher fidelities than the experiment and does not appear to capture the effects of the refocusing $\pi$ pulse, as shown in FIG.~\ref{fig:fbound}{\textbf{C}}. This suggests that the experimental system has slow drifts which can be refocused by $\pi$-pulses. Interestingly, it has been demonstrated that dynamical decoupling is remarkably effective at extending the lifetime of GHZ states~\cite{kaufmann17}.

We demonstrate in this work an experimentally scalable entanglement metric based on multiple quantum coherences, and applied it to verify 18-qubit multipartite GHZ entanglement. Our experiments show encouraging results in the ability to entangle and disentangle highly correlated many-body states in near term quantum devices.
We are exploring new variations of CX gates which can be applied simultaneously on adjacent qubit pairs while canceling $ZZ$ errors~\cite{Takita17}. This should improve gate fidelity in the entangling and disentangling steps. The lifetimes of MQC amplitudes should be measured and compared to those of parity oscillations, which were reported to decrease as $N^2$ in trapped ions~\cite{Monz2011} and $N$ in superconducting qubits~\cite{Ozaeta_2019}. It will be interesting to extend MQC to other entangled states such as the W-state and study their entanglement properties. Lastly, the newly developed error mitigation techniques~\cite{temme17l, Kandala2019} may give us insights to the maximum GHZ fidelity achievable in our device in the limit of zero noise.

During the preparation of this manuscript, we became aware of recent experiments demonstrating 18-qubit multipartite GHZ entanglement in a tunable-frequency transmon device~\cite{song19x} and 20-qubit multipartite GHZ entanglement in a Rydberg atoms array~\cite{omran19}.


\begin{acknowledgements}
We thank T. Alexander, L. Bishop, P. Cappellaro, J. M. Chow, A. C{\'o}rcoles, P. Jurcevic, A. Kandala, J. Kim, K. Krsulich, E. Magesan, S. Merkel, A. Mezzacapo, Z. Minev, P. Nation, J. Smolin, M. Steffen
, M. Takita, K. Temme, and C. Wood for insightful discussions. This work was supported by ARO under Contract No. W911NF-14-1-0124. The authors declare that they have no competing interests.
\end{acknowledgements}

\bibliographystyle{apsrev4-1}
\bibliography{qcvvbib}

\newpage
\onecolumngrid
\newpage

\input{sm}

\end{document}

%% file: circ1.tex
$$
\Qcircuit @C=0.7em @R=1.2em{
\raisebox{3em}{\bf{A}} &\lstick{\ket{0}} & \gate{H} & \multigate{3}{U_\text{CX}} & \multigate{3}{\text{Sensing}} & \multigate{3}{U^\dagger_\text{CX}} & \gate{H} & \meter \\ 
&\lstick{\ket{0}} & \qw & \ghost{U_\text{CX}}  &  \ghost{\text{Sensing}}  &\ghost{U^\dagger_\text{CX}} & \qw &\qw  \\
&\lstick{\vdots} & \qw & \ghost{U_\text{CX}} &  \ghost{\text{Sensing}}  &\ghost{U^\dagger_\text{CX}} & \qw &\qw  \\
&\lstick{\ket{0}} & \qw & \ghost{U_\text{CX}}  &  \ghost{\text{Sensing}}  &\ghost{U^\dagger_\text{CX}} & \qw &\qw 
}
$$



%% file: circ2.tex
$$
\Qcircuit @C=0.7em @R=0.75em{
\raisebox{3em}{\bf{B}} &\lstick{\ket{0}} & \gate{H} & \multigate{3}{U_\text{CX}} & \gate{U_\phi} & \multigate{3}{U^\dagger_\text{CX}} & \gate{H} & \meter \\
&\lstick{\ket{0}} & \qw & \ghost{U_\text{CX}}  &  \gate{U_\phi}  &\ghost{U^\dagger_\text{CX}} & \qw & \meter \\
&\lstick{\vdots} & \qw & \ghost{U_\text{CX}} &  \gate{U_\phi}  &\ghost{U^\dagger_\text{CX}} & \qw & \meter \\
&\lstick{\ket{0}} & \qw & \ghost{U_\text{CX}}  &  \gate{U_\phi}  &\ghost{U^\dagger_\text{CX}} & \qw & \meter
}
$$

%% file: circ3.tex
$$
\Qcircuit @C=0.7em @R=0.75em{
\raisebox{3em}{\bf{C}} & \lstick{\ket{0}} & \gate{H} & \multigate{3}{U_\text{CX}} & \gate{X} & \gate{U_\phi} & \multigate{3}{U^\dagger_\text{CX}} & \gate{H} & \meter \\
&\lstick{\ket{0}} & \qw & \ghost{U_\text{CX}} &  \gate{X} &  \gate{U_\phi}  &\ghost{U^\dagger_\text{CX}} & \qw & \meter \\
&\lstick{\vdots} & \qw & \ghost{U_\text{CX}} &  \gate{X} &  \gate{U_\phi}  &\ghost{U^\dagger_\text{CX}} & \qw & \meter \\
&\lstick{\ket{0}} & \qw & \ghost{U_\text{CX}} &  \gate{X} &  \gate{U_\phi}  &\ghost{U^\dagger_\text{CX}} & \qw & \meter
}
$$

%% file: circ4.tex
$$
\Qcircuit @C=0.75em @R=0.25em {
\raisebox{3em}{\bf{D}} &\lstick{\ket{0}_0}       &\qw &\qw 		&\qw &\targ &\ctrl{1} &\qw &\qw &\gate{X} &\gate{U_\phi} &\qw &\qw &\ctrl{1} &\targ &\qw &\qw &\qw & \meter \\
&\lstick{\ket{0}_1}      & \qw &\qw 		& \qw & \qw &\targ &\ctrl{1} &\qw &\gate{X} &\gate{U_\phi} &\qw &\ctrl{1} &\targ &\qw &\qw &\qw &\qw & \meter \\
&\lstick{\ket{0}_2}      & \qw &\qw 		& \qw & \qw &\qw &\targ &\ctrl{1} &\gate{X} &\gate{U_\phi} &\ctrl{1} &\targ &\qw &\qw &\qw &\qw &\qw & \meter \\
&\lstick{\ket{0}_3}      & \qw &\qw 		& \qw & \qw &\qw &\qw &\targ &\gate{X} &\gate{U_\phi} &\targ &\qw &\qw &\qw &\qw &\qw &\qw & \meter \\
&\lstick{\ket{0}_4}      & \qw &\qw 		& \qw & \qw &\qw &\qw &\targ &\gate{X} &\gate{U_\phi} &\targ &\qw &\qw &\qw &\qw &\qw &\qw & \meter \\
&\lstick{\ket{0}_5}      & \gate{H} &\ctrl{5} 	&\ctrl{1} &\ctrl{-5} &\qw & \qw  &\qw &\gate{X} &\gate{U_\phi} &\qw &\qw &\qw &\ctrl{-5} &\ctrl{1} &\ctrl{5} &\gate{H} & \meter \\
&\lstick{\ket{0}_6}      & \qw &\qw 		&\targ &\ctrl{1} &\qw &\qw &\qw &\gate{X} &\gate{U_\phi} &\qw &\qw &\qw &\ctrl{1} &\targ &\qw &\qw & \meter \\
&\lstick{\ket{0}_7}      & \qw &\qw 		& \qw &\targ &\ctrl{1} &\qw &\qw &\gate{X} &\gate{U_\phi} &\qw &\qw &\ctrl{1} &\targ &\qw &\qw &\qw & \meter \\
&\lstick{\ket{0}_8}      & \qw &\qw 		& \qw & \qw &\targ &\ctrl{1} &\qw &\gate{X} &\gate{U_\phi} &\qw &\ctrl{1} &\targ &\qw &\qw &\qw &\qw & \meter \\
&\lstick{\ket{0}_9}      & \qw &\qw 		& \qw & \qw &\qw &\targ &\ctrl{-5} &\gate{X} &\gate{U_\phi}  &\ctrl{-5} &\targ &\qw &\qw &\qw &\qw &\qw & \meter \\
&\lstick{\ket{0}_{10}} & \qw &\targ 	&\ctrl{1} &\qw &\ctrl{5}& \qw &\qw &\gate{X} &\gate{U_\phi} &\qw &\qw &\ctrl{5} &\qw &\ctrl{1} &\targ &\qw & \meter \\
&\lstick{\ket{0}_{11}} & \qw &\qw 		&\targ &\ctrl{1} & \qw & \qw &\qw &\gate{X} &\gate{U_\phi} &\qw &\qw &\qw &\ctrl{1} &\targ &\qw &\qw & \meter \\
&\lstick{\ket{0}_{12}} & \qw &\qw 		& \qw &\targ & \qw &\ctrl{1} &\qw &\gate{X} &\gate{U_\phi} &\qw &\ctrl{1} &\qw &\targ &\qw &\qw &\qw & \meter \\
&\lstick{\ket{0}_{13}} & \qw &\qw 		& \qw & \qw &\qw &\targ &\ctrl{1} &\gate{X} &\gate{U_\phi} &\ctrl{1} &\targ &\qw &\qw &\qw &\qw &\qw & \meter \\
&\lstick{\ket{0}_{14}} & \qw &\qw 		& \qw & \qw &\qw &\qw &\targ &\gate{X} &\gate{U_\phi} &\targ &\qw &\qw &\qw &\qw &\qw &\qw & \meter \\
&\lstick{\ket{0}_{15}} & \qw &\qw 		& \qw & \qw &\targ &\ctrl{1} &\qw &\gate{X} &\gate{U_\phi} &\qw &\ctrl{1} &\targ &\qw &\qw &\qw &\qw & \meter \\
&\lstick{\ket{0}_{16}} & \qw &\qw 		& \qw & \qw &\qw &\targ &\ctrl{1} &\gate{X} &\gate{U_\phi} &\ctrl{1} &\targ &\qw &\qw &\qw &\qw &\qw & \meter \\
&\lstick{\ket{0}_{17}} & \qw &\qw 		& \qw & \qw &\qw &\qw &\targ &\gate{X} &\gate{U_\phi} &\targ &\qw &\qw &\qw &\qw &\qw &\qw &\meter \\
}
$$

%% file: sm.tex
\section*{SUPPLEMENTARY MATERIAL}

\subsection{Multiple Quantum Coherences}
Consider writing the density matrix as $\rho=\sum_{m,m'} \rho_{m,m'} \ketbra{m}{m'}$, where the basis states satisfy $\sum_j \sigma_z^j/2 \ket{m} = m\ket{m}$. We can expand the density matrix as $\rho=\sum_q \rho_q$, where $\rho_q=\sum_m \rho_{m,m-q}\ketbra{m}{m-q}$. It can be shown that $\rho_q$ satisfies the following:
\begin{gather}
e^{-i\frac{\phi}{2}\sum_j \sigma_z^j} \rho_q e^{i\frac{\phi}{2}\sum_j \sigma_z^j} = e^{-i q \phi} \rho_q, \qquad \comm{\displaystyle{\sum}_j \sigma_z^j/2}{\rho_q}=q\rho_q
\label{eq:mqcprop}
\end{gather}
Since $\rho$ is hermitian we also have $\rho_q^\dagger = \rho_{-q}$. Each $\rho_q$ occupies a different part of the density matrix, and it obeys the orthogonality condition $\text{Tr}(\rho_q \rho_p)=\delta_{q,-p} \text{Tr}(\rho_q \rho_{-q})$. While each $\rho_q$ is not directly observable, the trace $I_q=\text{Tr}(\rho_q \rho_{-q})$ is. $I_q$ is the multiple quantum coherence amplitude, and it can be found by Fourier transforming the overlap signal $S_\phi = \text{Tr}(\rho_\phi \rho)$, where $\rho_\phi = e^{-i\frac{\phi}{2}\sum_j\sigma_z^j}\rho e^{i\frac{\phi}{2}\sum_j\sigma_z^j}$. For a general time-dependent density matrix, measuring $S_\phi$ requires the ability to implement time-reversed evolution. Upon expanding $\rho$ inside $S_\phi$ and using the first relation in Eq.~(\ref{eq:mqcprop}) we have
\begin{align*}
S_\phi&=\text{Tr}(\sum_q e^{-iq\phi} \rho_q \sum_p \rho_p) =\sum_q e^{-iq\phi} \text{Tr}(\rho_q \sum_p \rho_p) \\
	   &=\sum_q e^{-iq \phi}\text{Tr}(\rho_q \rho_{-q}) =\sum_q e^{-iq \phi} I_q
\end{align*}
where in the second to last step we used the orthogonality condition. Fourier transforming gives $I_q = \mathcal{N}^{-1}\sum_\phi e^{iq\phi}S_\phi$, where $\mathcal{N}$ is a normalization constant depending on the number of $\phi$ used in the experiments. If the maximum coherence order to be measured is $q_\text{max}$, we need at least $2q_\text{max}$ experiments. The angle $\phi$ can be chosen as $\phi=\frac{\pi j}{q_\text{max}}$ where $j=0,1,2,\cdots 2q_\text{max}-1$ and $\mathcal{N}=2 q_\text{max}$. While our discussion assumes unitary evolution, it has been shown that in certain types of decoherence the experimental method for measuring multiple quantum coherence is still valid~\cite{garttner18prl}.

For an ideal $N$-qubit GHZ state, the nonzero elements in the density matrix resides only in the four corners. Therefore only three components arise in the expansion: $\rho^\text{GHZ}=\rho_0^\text{GHZ} +\rho_N^\text{GHZ} +\rho_{-N}^\text{GHZ}$. Explicitly they are given by
\begin{align*}
\rho_0^\text{GHZ} & = \frac{1}{2}(\ketbra{000..00}{000..00}+\ketbra{111..11}{111..11}) \\
\rho_N^\text{GHZ} & = \frac{1}{2} \ketbra{000..00}{111..11} \\
\rho_{-N}^\text{GHZ} & = {\rho_N^\text{GHZ}}^\dagger
\end{align*}
the corresponding multiple quantum amplitudes are given by
\begin{align}\label{eq:ghzint}
I_0^\text{GHZ} = \text{Tr}(\rho_0^\text{GHZ} \rho_0^\text{GHZ}) = \frac{1}{2}, \qquad
I_N^\text{GHZ} = \text{Tr}(\rho_N^\text{GHZ} \rho_{-N}^\text{GHZ}) = \frac{1}{4}, \qquad
I_{-N}^\text{GHZ} = I_N^\text{GHZ}
\end{align}
Multiple quantum coherence amplitudes are symmetric: $I_q = I_{-q}$.

\subsection{Fidelity Bounds from MQC amplitudes}
The state fidelity, given by $F=\mel{\text{GHZ}}{\rho}{\text{GHZ}}=\text{Tr}(\rho \rho^\text{GHZ})$, can be bounded by:
\begin{align}\label{eq:fboundsm}
2\sqrt{I_N} \leq F \leq \sqrt{I_0/2} + \sqrt{I_N}
\end{align}
The upper bound on $F$ follows from the Cauchy-Schwarz inequality:
\begin{gather*}
\text{Tr}(\rho \rho^\text{GHZ}) =\sum_{q=-N}^N\text{Tr}(\rho_q \rho_{-q}^\text{GHZ}) \\
\leq \sum_{q=-N}^N \sqrt{\text{Tr}(\rho_q \rho_{-q})\text{Tr}(\rho_{q}^\text{GHZ}\rho_{-q}^\text{GHZ})} =\sqrt{I_0/2} + \sqrt{I_N}
\end{gather*}
To prove the lower bound we first notice that
\begin{align}\label{eq:flb}
F&=\text{Tr}(\rho_0 \rho_{0}^\text{GHZ})+\text{Tr}(\rho_N \rho_{-N}^\text{GHZ})+\text{Tr}(\rho_{-N} \rho_{N}^\text{GHZ}) \nonumber \\
&\geq 2(\text{Tr}(\rho_N \rho_{-N}^\text{GHZ})+\text{Tr}(\rho_{-N} \rho_{N}^\text{GHZ}))
\end{align}
This can be proved by writing the density matrix as $\rho=\sum_j w_j \ketbra{\psi_j}{\psi_j}$, where $w_j\geq 0$ and $\sum_j w_j=1$. The state vectors $\ket{\psi_j}$ need not to be orthogonal, we can in general expand $\ket{\psi_j}$ as
\begin{align*}
\ket{\psi_j}=\alpha_j \ket{00...0000} + \beta_j \ket{11...1111} +\cdots
\end{align*}
Since $|\alpha_j - \beta_j|^2\geq 0$, upon expanding we have $|\alpha_j|^2 + |\beta_j|^2 \geq \alpha_j \beta_j^* + \alpha_j^*\beta_j$. It then follows that
\begin{align*}
\frac{1}{2}\sum_j w_j (|\alpha_j|^2 + |\beta_j|^2) \geq \frac{1}{2}\sum_j w_j (\alpha_j \beta_j^* + \alpha_j^*\beta_j)
\end{align*}
which is the same as
\begin{align*}
\text{Tr}(\rho_0 \rho_{0}^\text{GHZ})\geq \text{Tr}(\rho_N \rho_{-N}^\text{GHZ})+\text{Tr}(\rho_{-N} \rho_{N}^\text{GHZ})
\end{align*}
thereby proving Eq.~(\ref{eq:flb}). One can go one step further by noting that $\rho_N = \kappa \rho_N^\text{GHZ}$, where $\kappa$ is a complex constant. Using $\text{Tr}(\rho_N \rho_{-N}^\text{GHZ}) = \kappa I_N^\text{GHZ} = \frac{1}{\kappa^*}I_N$ we can show that $|\kappa| = 2\sqrt{I_N}$. Notice $\kappa$ can always be made real by appropriately rotating the density matrix $\rho$. Substituting $\rho_N = 2\sqrt{I_N} \rho_N^\text{GHZ}$ into Eq.~(\ref{eq:flb}) gives $2\sqrt{I_N}$ as the lower bound of $F$. 
In addition to the fidelity bounds, we can also obtain the GHZ state fidelity $F$:
\begin{align*}
F=\frac{1}{2}(P_{000..00} + P_{111..11}) +\sqrt{I_N}
\end{align*} 
where $P_{000..00}=\expval{\rho}{000...00}$ and $P_{111..11}=\expval{\rho}{111...11}$ are the probabilities of finding all zeroes and all ones in the state $\rho$. 

\subsection{Readout Calibration}
\begin{figure*}
\centering
\includegraphics[width=0.98\textwidth]{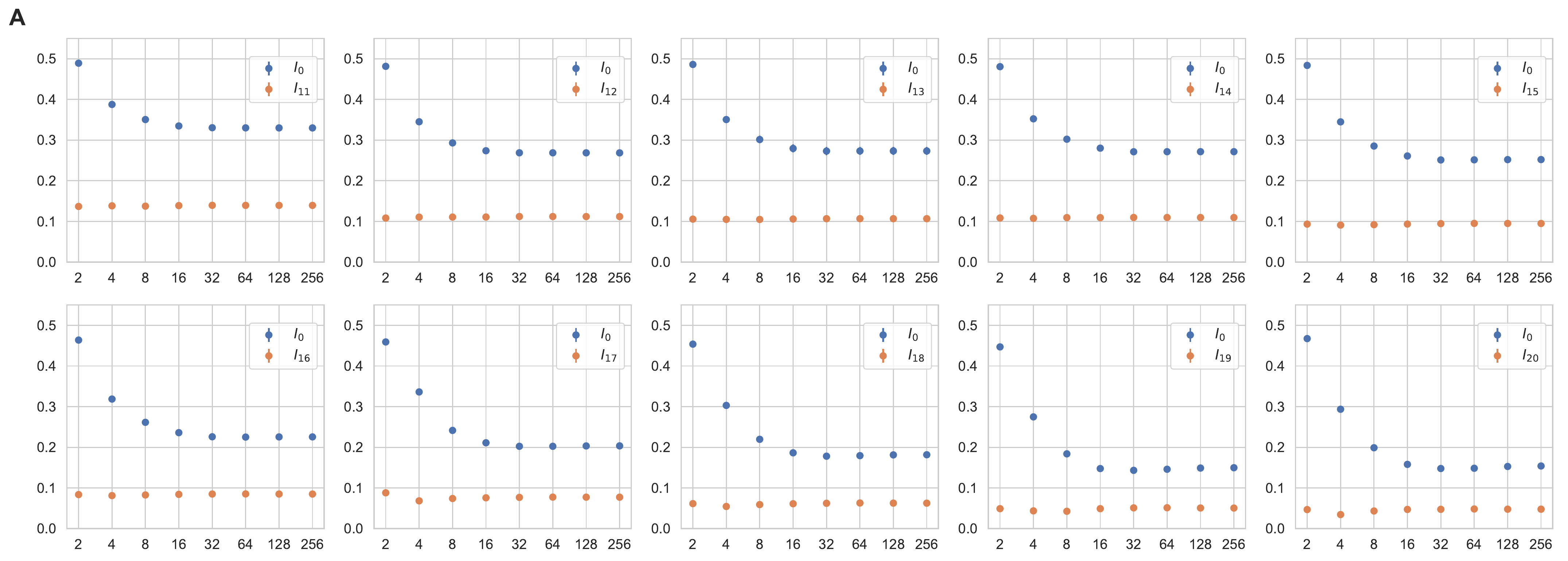} \\
\vspace{1cm}
\includegraphics[width=0.98\textwidth]{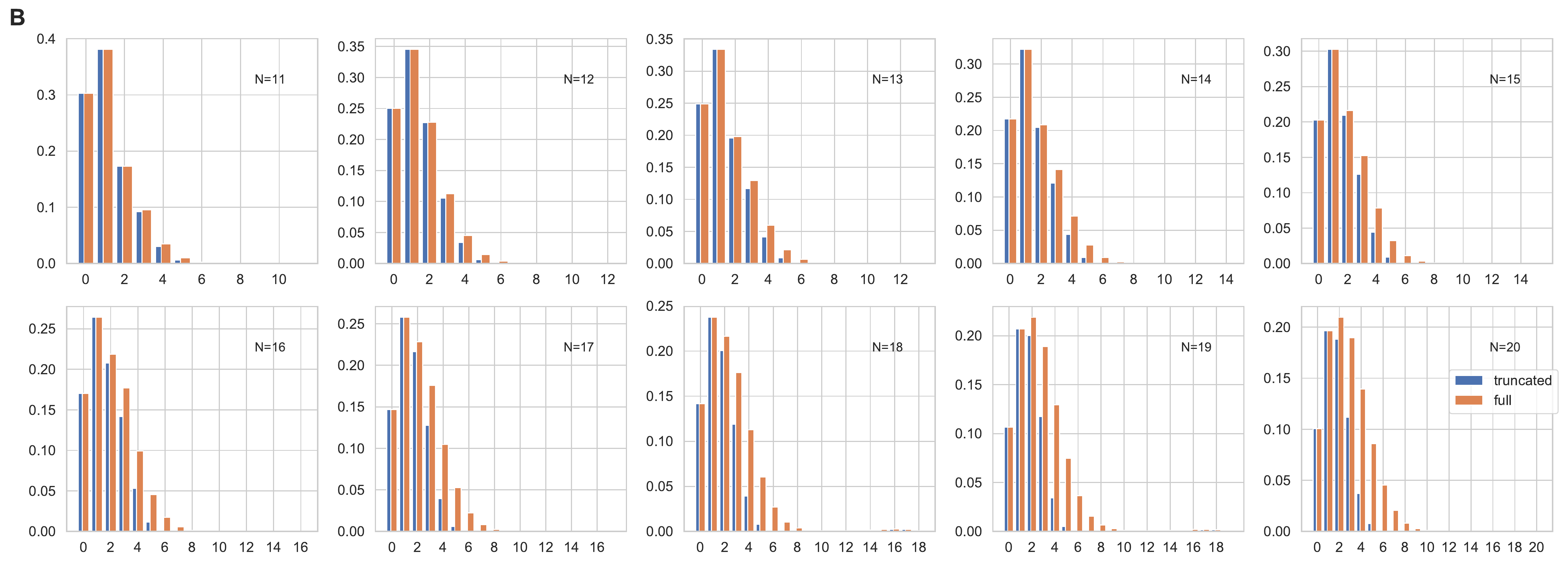} \\
\vspace{1cm}
\includegraphics[width=0.98\textwidth]{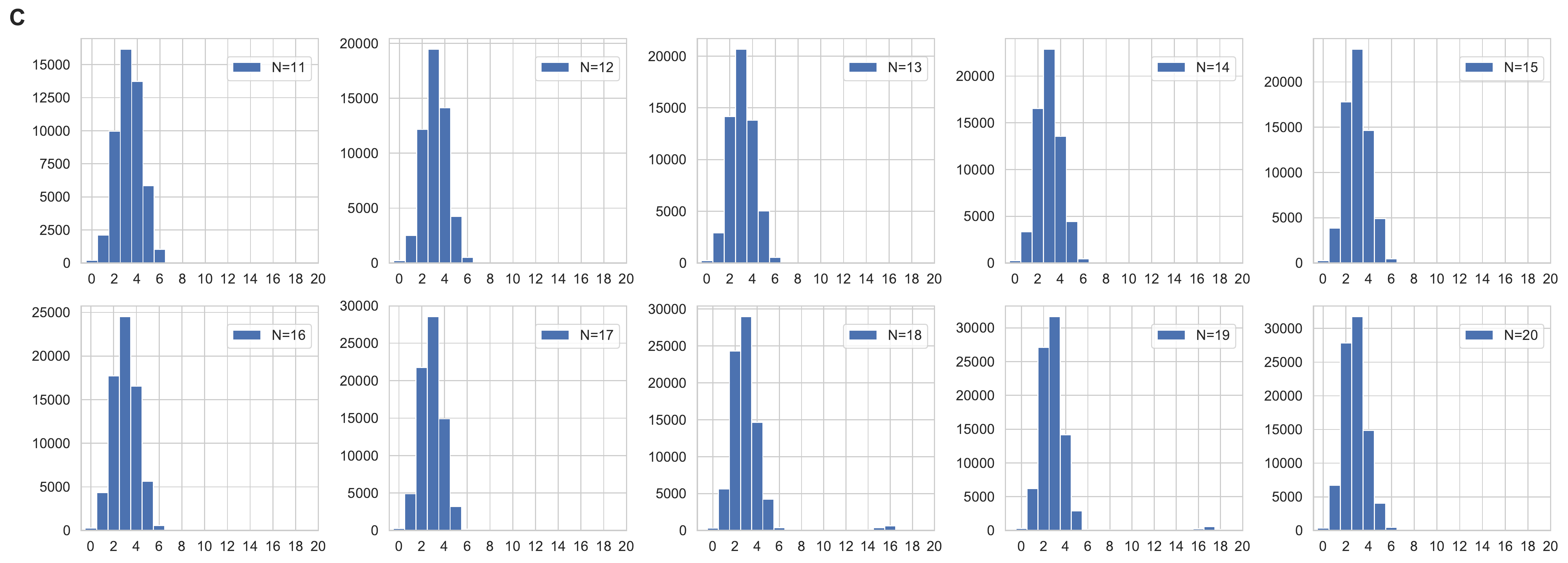}
\caption{\label{fig:alldat}\textbf{Additional Data for $\bf{N=11}$ to $\bf{N=20}$ MQC Experiments} {\textbf{A}}: readout corrected MQC amplitudes $I_0$ and $I_N$ as a function of the number of states used in $A_t$. {\textbf{B}}: the largest 256 counts and full counts for each $N$ are plotted as a function of excitation number. The full counts are normalized such that the sum over all excitation is one. {\textbf{C}}: the histogram of excitation number is plotted using the largest 256 counts from all experiments for each $N$. Interestingly, three excitation states have the highest probability for all $N$.}
\end{figure*}

We construct a truncated readout calibration matrix $A_t$ according to the largest 256 states by weight in the output of each experiments. In this section we justify using only 256 states as opposed to $2^N$ states. Our metric to verify GHZ entanglement requires MQC amplitudes $I_0$ and $I_N$. In FIG.~\ref{fig:alldat}{\textbf{A}} we plot the corrected $I_0$ and $I_N$ as a function of the number of states used in $A_t$; we see a rapid convergence after just 32 states. Interestingly, $I_0$ decreases as we add more states into the readout calibration, while $I_N$ is relatively unchanged. 

In FIG.~\ref{fig:alldat}{\textbf{B}} we compare all counts and the largest 256 counts from all experiments for each $N$. The counts are grouped according to excitation number (number of ones) and divided by the total number of counts. For small $N$, there is little difference between all counts and the largest 256 counts, indicating most of the weight in the output are contained in the largest 256 counts. As $N$ increases however, the total counts begins to spread out to higher excitation numbers, while the largest 256 counts are still localized in the low excitation numbers. This discrepancy however does not affect the calibrated values of the all zeroes count, as demonstrated by the convergence of MQC amplitudes.

In FIG.~\ref{fig:alldat}{\textbf{C}} we examine the distribution of the largest 245 counts as a function of excitation number for each $N$. Similar distributions centered around three excitation are observed for all $N$. 

\subsection{Device Parameters and Qubits involved in making GHZ states}
In Table~\ref{tab:device} we show the typical qubit parameters for the device. In Table~\ref{tab:qghz} we list the physical qubits involved in the state tomography experiments and the $N$-qubit MQC experiments on the device. The number labeling the physical qubit is shown in FIG. 1 in the main text.
\begin{table}[b]
\caption{\label{tab:device}\textbf{Qubit parameters on IBM System One} The qubit frequency, $T_1$, $T_{2,\text{echo}}$, readout fidelity are presented.}
\begin{ruledtabular}
\begin{tabular}{ccccc}
Qubit & Frequency (GHz) & $T_1$ ($\mu s$) & $T_{2,\text{echo}}$ ($\mu s$) & Readout Fidelity\\
\colrule
Q0 & 4.666 & 88.1 & 76.6 & 98.1 \\
Q1 & 4.760 & 69.0 & 75.7 & 96.4 \\
Q2 & 4.609 & 58.3 & 65.4 & 97.2 \\
Q3 & 5.031 & 60.9 & 73.0 & 79.7 \\
Q4 & 4.657 & 69.1 & 78.1 & 96.6 \\
Q5 & 4.752 & 74.4 & 71.9 & 95.9 \\
Q6 & 4.829 & 60.2 & 65.8 & 98.1 \\
Q7 & 4.698 & 80.7 & 79.5 & 96.4 \\
Q8 & 4.893 & 64.0 & 75.7 & 96.5 \\
Q9 & 4.731 & 63.3 & 70.7 & 93.0 \\
Q10 & 4.840 & 59.1 & 62.9 & 96.6 \\
Q11 & 4.755 & 64.1 & 56.3 & 97.8 \\
Q12 & 4.621 & 85.4 & 87.2 & 96.6 \\
Q13 & 4.859 & 69.4 & 83.2 & 93.6 \\
Q14 & 4.394 & 101.6 & 86.6 & 93.5 \\
Q15 & 4.693 & 76.1 & 74.3 & 98.1 \\
Q16 & 4.512 & 70.3 & 80.1 & 95.0 \\
Q17 & 4.719 & 66.4 & 79.2 & 97.8 \\
Q18 & 4.321 & 73.6 & 80.7 & 93.0 \\
Q19 & 4.593 & 83.3 & 85.5 & 97.6 \\
\colrule
Median & 4.708 & 69.2 & 76.2 & 96.6
\end{tabular}
\end{ruledtabular}
\end{table}

\begin{table}[b]
\caption{\label{tab:qghz}\textbf{Physical qubits used on IBM System One for state tomography and MQC experiments}}
\begin{ruledtabular}
\begin{tabular}{cl}
$N$ & Physical qubits used\\
\colrule
2 & [5, 10] \\
3 & [5, 10, 6] \\
4 & [5, 10, 6, 11] \\
5 & [5, 10, 6, 11, 0] \\
11 & [5, 10, 6, 11, 0, 12, 7, 15, 1, 8, 13] \\
12 & [5, 10, 6, 11, 0, 12, 7, 15, 1, 8, 13, 16] \\
13 & [5, 10, 6, 11, 0, 12, 7, 15, 1, 8, 13, 16, 2] \\
14 & [5, 10, 6, 11, 0, 12, 7, 15, 1, 8, 13, 16, 2, 9] \\
15 & [5, 10, 6, 11, 0, 12, 7, 15, 1, 8, 13, 16, 2, 9, 17] \\
16 & [5, 10, 6, 11, 0, 12, 7, 15, 1, 8, 13, 16, 2, 9, 17, 4] \\
17 & [5, 10, 6, 11, 0, 12, 7, 15, 1, 8, 13, 16, 2, 9, 17, 4, 14] \\
18 & [5, 10, 6, 11, 0, 12, 7, 15, 1, 8, 13, 16, 2, 9, 17, 4, 14, 3] \\
19 & [5, 10, 6, 11, 0, 12, 7, 15, 1, 8, 13, 16, 2, 9, 17, 4, 14, 3, 18] \\
20 & [5, 10, 6, 11, 0, 12, 7, 15, 1, 8, 13, 16, 2, 9, 17, 4, 14, 3, 18, 19] \\
\end{tabular}
\end{ruledtabular}
\end{table}

%% file: ghz_v2.bbl
\begin{thebibliography}{51}%
\makeatletter
\providecommand \@ifxundefined [1]{%
 \@ifx{#1\undefined}
}%
\providecommand \@ifnum [1]{%
 \ifnum #1\expandafter \@firstoftwo
 \else \expandafter \@secondoftwo
 \fi
}%
\providecommand \@ifx [1]{%
 \ifx #1\expandafter \@firstoftwo
 \else \expandafter \@secondoftwo
 \fi
}%
\providecommand \natexlab [1]{#1}%
\providecommand \enquote  [1]{``#1''}%
\providecommand \bibnamefont  [1]{#1}%
\providecommand \bibfnamefont [1]{#1}%
\providecommand \citenamefont [1]{#1}%
\providecommand \href@noop [0]{\@secondoftwo}%
\providecommand \href [0]{\begingroup \@sanitize@url \@href}%
\providecommand \@href[1]{\@@startlink{#1}\@@href}%
\providecommand \@@href[1]{\endgroup#1\@@endlink}%
\providecommand \@sanitize@url [0]{\catcode `\\12\catcode `\$12\catcode
  `\&12\catcode `\#12\catcode `\^12\catcode `\_12\catcode `\%12\relax}%
\providecommand \@@startlink[1]{}%
\providecommand \@@endlink[0]{}%
\providecommand \url  [0]{\begingroup\@sanitize@url \@url }%
\providecommand \@url [1]{\endgroup\@href {#1}{\urlprefix }}%
\providecommand \urlprefix  [0]{URL }%
\providecommand \Eprint [0]{\href }%
\providecommand \doibase [0]{http://dx.doi.org/}%
\providecommand \selectlanguage [0]{\@gobble}%
\providecommand \bibinfo  [0]{\@secondoftwo}%
\providecommand \bibfield  [0]{\@secondoftwo}%
\providecommand \translation [1]{[#1]}%
\providecommand \BibitemOpen [0]{}%
\providecommand \bibitemStop [0]{}%
\providecommand \bibitemNoStop [0]{.\EOS\space}%
\providecommand \EOS [0]{\spacefactor3000\relax}%
\providecommand \BibitemShut  [1]{\csname bibitem#1\endcsname}%
\let\auto@bib@innerbib\@empty
\bibitem [{\citenamefont {Shor}(1996)}]{shor96}%
  \BibitemOpen
  \bibfield  {author} {\bibinfo {author} {\bibfnamefont {P.}~\bibnamefont
  {Shor}},\ }\href@noop {} {\bibfield  {journal} {\bibinfo  {journal} {Proc.
  37nd Annual Symposium on Foundations of Computer Science}\ ,\ \bibinfo
  {pages} {56}} (\bibinfo {year} {1996})}\BibitemShut {NoStop}%
\bibitem [{\citenamefont {Bravyi}\ \emph {et~al.}(2018)\citenamefont {Bravyi},
  \citenamefont {Gosset},\ and\ \citenamefont {K{\"o}nig}}]{Bravyi308}%
  \BibitemOpen
  \bibfield  {author} {\bibinfo {author} {\bibfnamefont {S.}~\bibnamefont
  {Bravyi}}, \bibinfo {author} {\bibfnamefont {D.}~\bibnamefont {Gosset}}, \
  and\ \bibinfo {author} {\bibfnamefont {R.}~\bibnamefont {K{\"o}nig}},\ }\href
  {\doibase 10.1126/science.aar3106} {\bibfield  {journal} {\bibinfo  {journal}
  {Science}\ }\textbf {\bibinfo {volume} {362}},\ \bibinfo {pages} {308}
  (\bibinfo {year} {2018})}\BibitemShut {NoStop}%
\bibitem [{\citenamefont {Preskill}(2018)}]{Preskill2018}%
  \BibitemOpen
  \bibfield  {author} {\bibinfo {author} {\bibfnamefont {J.}~\bibnamefont
  {Preskill}},\ }\href {\doibase 10.22331/q-2018-08-06-79} {\bibfield
  {journal} {\bibinfo  {journal} {{Quantum}}\ }\textbf {\bibinfo {volume}
  {2}},\ \bibinfo {pages} {79} (\bibinfo {year} {2018})}\BibitemShut {NoStop}%
\bibitem [{\citenamefont {Knill}\ \emph {et~al.}(2008)\citenamefont {Knill},
  \citenamefont {Leibfried}, \citenamefont {Reichle}, \citenamefont {Britton},
  \citenamefont {Blakestad}, \citenamefont {Jost}, \citenamefont {Langer},
  \citenamefont {Ozeri}, \citenamefont {Seidelin},\ and\ \citenamefont
  {Wineland}}]{knill2008}%
  \BibitemOpen
  \bibfield  {author} {\bibinfo {author} {\bibfnamefont {E.}~\bibnamefont
  {Knill}}, \bibinfo {author} {\bibfnamefont {D.}~\bibnamefont {Leibfried}},
  \bibinfo {author} {\bibfnamefont {R.}~\bibnamefont {Reichle}}, \bibinfo
  {author} {\bibfnamefont {J.}~\bibnamefont {Britton}}, \bibinfo {author}
  {\bibfnamefont {R.~B.}\ \bibnamefont {Blakestad}}, \bibinfo {author}
  {\bibfnamefont {J.~D.}\ \bibnamefont {Jost}}, \bibinfo {author}
  {\bibfnamefont {C.}~\bibnamefont {Langer}}, \bibinfo {author} {\bibfnamefont
  {R.}~\bibnamefont {Ozeri}}, \bibinfo {author} {\bibfnamefont
  {S.}~\bibnamefont {Seidelin}}, \ and\ \bibinfo {author} {\bibfnamefont
  {D.~J.}\ \bibnamefont {Wineland}},\ }\href {\doibase
  10.1103/PhysRevA.77.012307} {\bibfield  {journal} {\bibinfo  {journal} {Phys.
  Rev. A}\ }\textbf {\bibinfo {volume} {77}},\ \bibinfo {pages} {012307}
  (\bibinfo {year} {2008})}\BibitemShut {NoStop}%
\bibitem [{\citenamefont {Magesan}\ \emph
  {et~al.}(2012{\natexlab{a}})\citenamefont {Magesan}, \citenamefont
  {Gambetta},\ and\ \citenamefont {Emerson}}]{Magesan2012}%
  \BibitemOpen
  \bibfield  {author} {\bibinfo {author} {\bibfnamefont {E.}~\bibnamefont
  {Magesan}}, \bibinfo {author} {\bibfnamefont {J.~M.}\ \bibnamefont
  {Gambetta}}, \ and\ \bibinfo {author} {\bibfnamefont {J.}~\bibnamefont
  {Emerson}},\ }\href {\doibase 10.1103/PhysRevA.85.042311} {\bibfield
  {journal} {\bibinfo  {journal} {Phys. Rev. A}\ }\textbf {\bibinfo {volume}
  {85}},\ \bibinfo {pages} {042311} (\bibinfo {year}
  {2012}{\natexlab{a}})}\BibitemShut {NoStop}%
\bibitem [{\citenamefont {Magesan}\ \emph
  {et~al.}(2012{\natexlab{b}})\citenamefont {Magesan}, \citenamefont
  {Gambetta}, \citenamefont {Johnson}, \citenamefont {Ryan}, \citenamefont
  {Chow}, \citenamefont {Merkel}, \citenamefont {da~Silva}, \citenamefont
  {Keefe}, \citenamefont {Rothwell}, \citenamefont {Ohki}, \citenamefont
  {Ketchen},\ and\ \citenamefont {Steffen}}]{Magesan2012a}%
  \BibitemOpen
  \bibfield  {author} {\bibinfo {author} {\bibfnamefont {E.}~\bibnamefont
  {Magesan}}, \bibinfo {author} {\bibfnamefont {J.~M.}\ \bibnamefont
  {Gambetta}}, \bibinfo {author} {\bibfnamefont {B.~R.}\ \bibnamefont
  {Johnson}}, \bibinfo {author} {\bibfnamefont {C.~A.}\ \bibnamefont {Ryan}},
  \bibinfo {author} {\bibfnamefont {J.~M.}\ \bibnamefont {Chow}}, \bibinfo
  {author} {\bibfnamefont {S.~T.}\ \bibnamefont {Merkel}}, \bibinfo {author}
  {\bibfnamefont {M.~P.}\ \bibnamefont {da~Silva}}, \bibinfo {author}
  {\bibfnamefont {G.~A.}\ \bibnamefont {Keefe}}, \bibinfo {author}
  {\bibfnamefont {M.~B.}\ \bibnamefont {Rothwell}}, \bibinfo {author}
  {\bibfnamefont {T.~A.}\ \bibnamefont {Ohki}}, \bibinfo {author}
  {\bibfnamefont {M.~B.}\ \bibnamefont {Ketchen}}, \ and\ \bibinfo {author}
  {\bibfnamefont {M.}~\bibnamefont {Steffen}},\ }\href {\doibase
  10.1103/PhysRevLett.109.080505} {\bibfield  {journal} {\bibinfo  {journal}
  {Phys. Rev. Lett.}\ }\textbf {\bibinfo {volume} {109}},\ \bibinfo {pages}
  {080505} (\bibinfo {year} {2012}{\natexlab{b}})}\BibitemShut {NoStop}%
\bibitem [{\citenamefont {Sheldon}\ \emph
  {et~al.}(2016{\natexlab{a}})\citenamefont {Sheldon}, \citenamefont {Bishop},
  \citenamefont {Magesan}, \citenamefont {Filipp}, \citenamefont {Chow},\ and\
  \citenamefont {Gambetta}}]{sheldonIRB}%
  \BibitemOpen
  \bibfield  {author} {\bibinfo {author} {\bibfnamefont {S.}~\bibnamefont
  {Sheldon}}, \bibinfo {author} {\bibfnamefont {L.~S.}\ \bibnamefont {Bishop}},
  \bibinfo {author} {\bibfnamefont {E.}~\bibnamefont {Magesan}}, \bibinfo
  {author} {\bibfnamefont {S.}~\bibnamefont {Filipp}}, \bibinfo {author}
  {\bibfnamefont {J.~M.}\ \bibnamefont {Chow}}, \ and\ \bibinfo {author}
  {\bibfnamefont {J.~M.}\ \bibnamefont {Gambetta}},\ }\href {\doibase
  10.1103/PhysRevA.93.012301} {\bibfield  {journal} {\bibinfo  {journal} {Phys.
  Rev. A}\ }\textbf {\bibinfo {volume} {93}},\ \bibinfo {pages} {012301}
  (\bibinfo {year} {2016}{\natexlab{a}})}\BibitemShut {NoStop}%
\bibitem [{\citenamefont {Flammia}\ and\ \citenamefont {Liu}(2011)}]{FL}%
  \BibitemOpen
  \bibfield  {author} {\bibinfo {author} {\bibfnamefont {S.}~\bibnamefont
  {Flammia}}\ and\ \bibinfo {author} {\bibfnamefont {Y.-K.}\ \bibnamefont
  {Liu}},\ }\href {\doibase 10.1103/PhysRevLett.106.230501} {\bibfield
  {journal} {\bibinfo  {journal} {Phys. Rev. Lett.}\ }\textbf {\bibinfo
  {volume} {106}},\ \bibinfo {pages} {230501} (\bibinfo {year}
  {2011})}\BibitemShut {NoStop}%
\bibitem [{\citenamefont {da~Silva}\ \emph {et~al.}(2011)\citenamefont
  {da~Silva}, \citenamefont {Landon-Cardinal},\ and\ \citenamefont
  {Poulin}}]{SLCP}%
  \BibitemOpen
  \bibfield  {author} {\bibinfo {author} {\bibfnamefont {M.~P.}\ \bibnamefont
  {da~Silva}}, \bibinfo {author} {\bibfnamefont {O.}~\bibnamefont
  {Landon-Cardinal}}, \ and\ \bibinfo {author} {\bibfnamefont {D.}~\bibnamefont
  {Poulin}},\ }\href {\doibase 10.1103/PhysRevLett.107.210404} {\bibfield
  {journal} {\bibinfo  {journal} {Phys. Rev. Lett.}\ }\textbf {\bibinfo
  {volume} {107}},\ \bibinfo {pages} {210404} (\bibinfo {year}
  {2011})}\BibitemShut {NoStop}%
\bibitem [{\citenamefont {McKay}\ \emph
  {et~al.}(2017{\natexlab{a}})\citenamefont {McKay}, \citenamefont {Sheldon},
  \citenamefont {Smolin}, \citenamefont {Chow},\ and\ \citenamefont
  {Gambetta}}]{McKay17}%
  \BibitemOpen
  \bibfield  {author} {\bibinfo {author} {\bibfnamefont {D.~C.}\ \bibnamefont
  {McKay}}, \bibinfo {author} {\bibfnamefont {S.}~\bibnamefont {Sheldon}},
  \bibinfo {author} {\bibfnamefont {J.~A.}\ \bibnamefont {Smolin}}, \bibinfo
  {author} {\bibfnamefont {J.~M.}\ \bibnamefont {Chow}}, \ and\ \bibinfo
  {author} {\bibfnamefont {J.~M.}\ \bibnamefont {Gambetta}},\ }\href
  {https://arxiv.org/abs/1712.06550} {\bibfield  {journal} {\bibinfo  {journal}
  {arXiv preprint}\ }\textbf {\bibinfo {volume} {arXiv:1712.06550}} (\bibinfo
  {year} {2017}{\natexlab{a}})}\BibitemShut {NoStop}%
\bibitem [{\citenamefont {{Proctor}}\ \emph {et~al.}(2018)\citenamefont
  {{Proctor}}, \citenamefont {{Carignan-Dugas}}, \citenamefont {{Rudinger}},
  \citenamefont {{Nielsen}}, \citenamefont {{Blume-Kohout}},\ and\
  \citenamefont {{Young}}}]{proctor18}%
  \BibitemOpen
  \bibfield  {author} {\bibinfo {author} {\bibfnamefont {T.~J.}\ \bibnamefont
  {{Proctor}}}, \bibinfo {author} {\bibfnamefont {A.}~\bibnamefont
  {{Carignan-Dugas}}}, \bibinfo {author} {\bibfnamefont {K.}~\bibnamefont
  {{Rudinger}}}, \bibinfo {author} {\bibfnamefont {E.}~\bibnamefont
  {{Nielsen}}}, \bibinfo {author} {\bibfnamefont {R.}~\bibnamefont
  {{Blume-Kohout}}}, \ and\ \bibinfo {author} {\bibfnamefont {K.}~\bibnamefont
  {{Young}}},\ }\href@noop {} {\bibfield  {journal} {\bibinfo  {journal} {arXiv
  e-prints}\ } (\bibinfo {year} {2018})},\ \Eprint
  {http://arxiv.org/abs/1807.07975} {arXiv:1807.07975 [quant-ph]} \BibitemShut
  {NoStop}%
\bibitem [{\citenamefont {{Cross}}\ \emph {et~al.}(2018)\citenamefont
  {{Cross}}, \citenamefont {{Bishop}}, \citenamefont {{Sheldon}}, \citenamefont
  {{Nation}},\ and\ \citenamefont {{Gambetta}}}]{Cross2018}%
  \BibitemOpen
  \bibfield  {author} {\bibinfo {author} {\bibfnamefont {A.~W.}\ \bibnamefont
  {{Cross}}}, \bibinfo {author} {\bibfnamefont {L.~S.}\ \bibnamefont
  {{Bishop}}}, \bibinfo {author} {\bibfnamefont {S.}~\bibnamefont {{Sheldon}}},
  \bibinfo {author} {\bibfnamefont {P.~D.}\ \bibnamefont {{Nation}}}, \ and\
  \bibinfo {author} {\bibfnamefont {J.~M.}\ \bibnamefont {{Gambetta}}},\
  }\href@noop {} {\bibfield  {journal} {\bibinfo  {journal} {arXiv e-prints}\ }
  (\bibinfo {year} {2018})},\ \Eprint {http://arxiv.org/abs/1811.12926}
  {arXiv:1811.12926 [quant-ph]} \BibitemShut {NoStop}%
\bibitem [{\citenamefont {Wang}\ \emph
  {et~al.}(2018{\natexlab{a}})\citenamefont {Wang}, \citenamefont {Li},
  \citenamefont {Yin},\ and\ \citenamefont {Zeng}}]{Wang2018}%
  \BibitemOpen
  \bibfield  {author} {\bibinfo {author} {\bibfnamefont {Y.}~\bibnamefont
  {Wang}}, \bibinfo {author} {\bibfnamefont {Y.}~\bibnamefont {Li}}, \bibinfo
  {author} {\bibfnamefont {Z.-q.}\ \bibnamefont {Yin}}, \ and\ \bibinfo
  {author} {\bibfnamefont {B.}~\bibnamefont {Zeng}},\ }\href {\doibase
  10.1038/s41534-018-0095-x} {\bibfield  {journal} {\bibinfo  {journal} {npj
  Quantum Information}\ }\textbf {\bibinfo {volume} {4}},\ \bibinfo {pages}
  {46} (\bibinfo {year} {2018}{\natexlab{a}})}\BibitemShut {NoStop}%
\bibitem [{\citenamefont {{Mooney}}\ \emph {et~al.}(2019)\citenamefont
  {{Mooney}}, \citenamefont {{Hill}},\ and\ \citenamefont
  {{Hollenberg}}}]{mooney19x}%
  \BibitemOpen
  \bibfield  {author} {\bibinfo {author} {\bibfnamefont {G.~J.}\ \bibnamefont
  {{Mooney}}}, \bibinfo {author} {\bibfnamefont {C.~D.}\ \bibnamefont
  {{Hill}}}, \ and\ \bibinfo {author} {\bibfnamefont {L.~C.~L.}\ \bibnamefont
  {{Hollenberg}}},\ }\href@noop {} {\bibfield  {journal} {\bibinfo  {journal}
  {arXiv e-prints}\ } (\bibinfo {year} {2019})},\ \Eprint
  {http://arxiv.org/abs/1903.11747} {arXiv:1903.11747 [quant-ph]} \BibitemShut
  {NoStop}%
\bibitem [{\citenamefont {Shor}(1997)}]{shor97}%
  \BibitemOpen
  \bibfield  {author} {\bibinfo {author} {\bibfnamefont {P.}~\bibnamefont
  {Shor}},\ }\href {\doibase 10.1137/S0097539795293172} {\bibfield  {journal}
  {\bibinfo  {journal} {SIAM Journal on Computing}\ }\textbf {\bibinfo {volume}
  {26}},\ \bibinfo {pages} {1484} (\bibinfo {year} {1997})},\ \Eprint
  {http://arxiv.org/abs/https://doi.org/10.1137/S0097539795293172}
  {https://doi.org/10.1137/S0097539795293172} \BibitemShut {NoStop}%
\bibitem [{\citenamefont {Raussendorf}\ and\ \citenamefont
  {Briegel}(2001)}]{raussendorf01l}%
  \BibitemOpen
  \bibfield  {author} {\bibinfo {author} {\bibfnamefont {R.}~\bibnamefont
  {Raussendorf}}\ and\ \bibinfo {author} {\bibfnamefont {H.~J.}\ \bibnamefont
  {Briegel}},\ }\href {\doibase 10.1103/PhysRevLett.86.5188} {\bibfield
  {journal} {\bibinfo  {journal} {Phys. Rev. Lett.}\ }\textbf {\bibinfo
  {volume} {86}},\ \bibinfo {pages} {5188} (\bibinfo {year}
  {2001})}\BibitemShut {NoStop}%
\bibitem [{\citenamefont {Song}\ \emph {et~al.}(2017)\citenamefont {Song},
  \citenamefont {Xu}, \citenamefont {Liu}, \citenamefont {ping Yang},
  \citenamefont {Zheng}, \citenamefont {Deng}, \citenamefont {Xie},
  \citenamefont {Huang}, \citenamefont {Guo}, \citenamefont {Zhang},
  \citenamefont {Zhang}, \citenamefont {Xu}, \citenamefont {Zheng},
  \citenamefont {Zhu}, \citenamefont {Wang}, \citenamefont {Chen},
  \citenamefont {Lu}, \citenamefont {Han},\ and\ \citenamefont {Pan}}]{Song17}%
  \BibitemOpen
  \bibfield  {author} {\bibinfo {author} {\bibfnamefont {C.}~\bibnamefont
  {Song}}, \bibinfo {author} {\bibfnamefont {K.}~\bibnamefont {Xu}}, \bibinfo
  {author} {\bibfnamefont {W.}~\bibnamefont {Liu}}, \bibinfo {author}
  {\bibfnamefont {C.}~\bibnamefont {ping Yang}}, \bibinfo {author}
  {\bibfnamefont {S.-B.}\ \bibnamefont {Zheng}}, \bibinfo {author}
  {\bibfnamefont {H.}~\bibnamefont {Deng}}, \bibinfo {author} {\bibfnamefont
  {Q.}~\bibnamefont {Xie}}, \bibinfo {author} {\bibfnamefont {K.}~\bibnamefont
  {Huang}}, \bibinfo {author} {\bibfnamefont {Q.}~\bibnamefont {Guo}}, \bibinfo
  {author} {\bibfnamefont {L.}~\bibnamefont {Zhang}}, \bibinfo {author}
  {\bibfnamefont {P.}~\bibnamefont {Zhang}}, \bibinfo {author} {\bibfnamefont
  {D.}~\bibnamefont {Xu}}, \bibinfo {author} {\bibfnamefont {D.}~\bibnamefont
  {Zheng}}, \bibinfo {author} {\bibfnamefont {X.}~\bibnamefont {Zhu}}, \bibinfo
  {author} {\bibfnamefont {H.}~\bibnamefont {Wang}}, \bibinfo {author}
  {\bibfnamefont {Y.-A.}\ \bibnamefont {Chen}}, \bibinfo {author}
  {\bibfnamefont {C.-Y.}\ \bibnamefont {Lu}}, \bibinfo {author} {\bibfnamefont
  {S.}~\bibnamefont {Han}}, \ and\ \bibinfo {author} {\bibfnamefont {J.-W.}\
  \bibnamefont {Pan}},\ }\href {\doibase 10.1103/PhysRevLett.119.180511}
  {\bibfield  {journal} {\bibinfo  {journal} {Phys. Rev. Lett.}\ }\textbf
  {\bibinfo {volume} {119}},\ \bibinfo {pages} {180511} (\bibinfo {year}
  {2017})}\BibitemShut {NoStop}%
\bibitem [{\citenamefont {Monz}\ \emph {et~al.}(2011)\citenamefont {Monz},
  \citenamefont {Schindler}, \citenamefont {Barreiro}, \citenamefont {Chwalla},
  \citenamefont {Nigg}, \citenamefont {Coish}, \citenamefont {Harlander},
  \citenamefont {H\"ansel}, \citenamefont {Hennrich},\ and\ \citenamefont
  {Blatt}}]{Monz2011}%
  \BibitemOpen
  \bibfield  {author} {\bibinfo {author} {\bibfnamefont {T.}~\bibnamefont
  {Monz}}, \bibinfo {author} {\bibfnamefont {P.}~\bibnamefont {Schindler}},
  \bibinfo {author} {\bibfnamefont {J.~T.}\ \bibnamefont {Barreiro}}, \bibinfo
  {author} {\bibfnamefont {M.}~\bibnamefont {Chwalla}}, \bibinfo {author}
  {\bibfnamefont {D.}~\bibnamefont {Nigg}}, \bibinfo {author} {\bibfnamefont
  {W.~A.}\ \bibnamefont {Coish}}, \bibinfo {author} {\bibfnamefont
  {M.}~\bibnamefont {Harlander}}, \bibinfo {author} {\bibfnamefont
  {W.}~\bibnamefont {H\"ansel}}, \bibinfo {author} {\bibfnamefont
  {M.}~\bibnamefont {Hennrich}}, \ and\ \bibinfo {author} {\bibfnamefont
  {R.}~\bibnamefont {Blatt}},\ }\href {\doibase 10.1103/PhysRevLett.106.130506}
  {\bibfield  {journal} {\bibinfo  {journal} {Phys. Rev. Lett.}\ }\textbf
  {\bibinfo {volume} {106}},\ \bibinfo {pages} {130506} (\bibinfo {year}
  {2011})}\BibitemShut {NoStop}%
\bibitem [{\citenamefont {Wang}\ \emph
  {et~al.}(2018{\natexlab{b}})\citenamefont {Wang}, \citenamefont {Luo},
  \citenamefont {Huang}, \citenamefont {Chen}, \citenamefont {Su},
  \citenamefont {Liu}, \citenamefont {Chen}, \citenamefont {Li}, \citenamefont
  {Fang}, \citenamefont {Jiang}, \citenamefont {Zhang}, \citenamefont {Li},
  \citenamefont {Liu}, \citenamefont {Lu},\ and\ \citenamefont
  {Pan}}]{wang18l}%
  \BibitemOpen
  \bibfield  {author} {\bibinfo {author} {\bibfnamefont {X.-L.}\ \bibnamefont
  {Wang}}, \bibinfo {author} {\bibfnamefont {Y.-H.}\ \bibnamefont {Luo}},
  \bibinfo {author} {\bibfnamefont {H.-L.}\ \bibnamefont {Huang}}, \bibinfo
  {author} {\bibfnamefont {M.-C.}\ \bibnamefont {Chen}}, \bibinfo {author}
  {\bibfnamefont {Z.-E.}\ \bibnamefont {Su}}, \bibinfo {author} {\bibfnamefont
  {C.}~\bibnamefont {Liu}}, \bibinfo {author} {\bibfnamefont {C.}~\bibnamefont
  {Chen}}, \bibinfo {author} {\bibfnamefont {W.}~\bibnamefont {Li}}, \bibinfo
  {author} {\bibfnamefont {Y.-Q.}\ \bibnamefont {Fang}}, \bibinfo {author}
  {\bibfnamefont {X.}~\bibnamefont {Jiang}}, \bibinfo {author} {\bibfnamefont
  {J.}~\bibnamefont {Zhang}}, \bibinfo {author} {\bibfnamefont
  {L.}~\bibnamefont {Li}}, \bibinfo {author} {\bibfnamefont {N.-L.}\
  \bibnamefont {Liu}}, \bibinfo {author} {\bibfnamefont {C.-Y.}\ \bibnamefont
  {Lu}}, \ and\ \bibinfo {author} {\bibfnamefont {J.-W.}\ \bibnamefont {Pan}},\
  }\href {\doibase 10.1103/PhysRevLett.120.260502} {\bibfield  {journal}
  {\bibinfo  {journal} {Phys. Rev. Lett.}\ }\textbf {\bibinfo {volume} {120}},\
  \bibinfo {pages} {260502} (\bibinfo {year} {2018}{\natexlab{b}})}\BibitemShut
  {NoStop}%
\bibitem [{\citenamefont {Gong}\ \emph {et~al.}(2019)\citenamefont {Gong},
  \citenamefont {Chen}, \citenamefont {Zheng}, \citenamefont {Wang},
  \citenamefont {Zha}, \citenamefont {Deng}, \citenamefont {Yan}, \citenamefont
  {Rong}, \citenamefont {Wu}, \citenamefont {Li}, \citenamefont {Chen},
  \citenamefont {Zhao}, \citenamefont {Liang}, \citenamefont {Lin},
  \citenamefont {Xu}, \citenamefont {Guo}, \citenamefont {Sun}, \citenamefont
  {Castellano}, \citenamefont {Wang}, \citenamefont {Peng}, \citenamefont {Lu},
  \citenamefont {Zhu},\ and\ \citenamefont {Pan}}]{gong19l}%
  \BibitemOpen
  \bibfield  {author} {\bibinfo {author} {\bibfnamefont {M.}~\bibnamefont
  {Gong}}, \bibinfo {author} {\bibfnamefont {M.-C.}\ \bibnamefont {Chen}},
  \bibinfo {author} {\bibfnamefont {Y.}~\bibnamefont {Zheng}}, \bibinfo
  {author} {\bibfnamefont {S.}~\bibnamefont {Wang}}, \bibinfo {author}
  {\bibfnamefont {C.}~\bibnamefont {Zha}}, \bibinfo {author} {\bibfnamefont
  {H.}~\bibnamefont {Deng}}, \bibinfo {author} {\bibfnamefont {Z.}~\bibnamefont
  {Yan}}, \bibinfo {author} {\bibfnamefont {H.}~\bibnamefont {Rong}}, \bibinfo
  {author} {\bibfnamefont {Y.}~\bibnamefont {Wu}}, \bibinfo {author}
  {\bibfnamefont {S.}~\bibnamefont {Li}}, \bibinfo {author} {\bibfnamefont
  {F.}~\bibnamefont {Chen}}, \bibinfo {author} {\bibfnamefont {Y.}~\bibnamefont
  {Zhao}}, \bibinfo {author} {\bibfnamefont {F.}~\bibnamefont {Liang}},
  \bibinfo {author} {\bibfnamefont {J.}~\bibnamefont {Lin}}, \bibinfo {author}
  {\bibfnamefont {Y.}~\bibnamefont {Xu}}, \bibinfo {author} {\bibfnamefont
  {C.}~\bibnamefont {Guo}}, \bibinfo {author} {\bibfnamefont {L.}~\bibnamefont
  {Sun}}, \bibinfo {author} {\bibfnamefont {A.~D.}\ \bibnamefont {Castellano}},
  \bibinfo {author} {\bibfnamefont {H.}~\bibnamefont {Wang}}, \bibinfo {author}
  {\bibfnamefont {C.}~\bibnamefont {Peng}}, \bibinfo {author} {\bibfnamefont
  {C.-Y.}\ \bibnamefont {Lu}}, \bibinfo {author} {\bibfnamefont
  {X.}~\bibnamefont {Zhu}}, \ and\ \bibinfo {author} {\bibfnamefont {J.-W.}\
  \bibnamefont {Pan}},\ }\href {\doibase 10.1103/PhysRevLett.122.110501}
  {\bibfield  {journal} {\bibinfo  {journal} {Phys. Rev. Lett.}\ }\textbf
  {\bibinfo {volume} {122}},\ \bibinfo {pages} {110501} (\bibinfo {year}
  {2019})}\BibitemShut {NoStop}%
\bibitem [{\citenamefont {Degen}\ \emph {et~al.}(2017)\citenamefont {Degen},
  \citenamefont {Reinhard},\ and\ \citenamefont {Cappellaro}}]{degen17rmp}%
  \BibitemOpen
  \bibfield  {author} {\bibinfo {author} {\bibfnamefont {C.~L.}\ \bibnamefont
  {Degen}}, \bibinfo {author} {\bibfnamefont {F.}~\bibnamefont {Reinhard}}, \
  and\ \bibinfo {author} {\bibfnamefont {P.}~\bibnamefont {Cappellaro}},\
  }\href {\doibase 10.1103/RevModPhys.89.035002} {\bibfield  {journal}
  {\bibinfo  {journal} {Rev. Mod. Phys.}\ }\textbf {\bibinfo {volume} {89}},\
  \bibinfo {pages} {035002} (\bibinfo {year} {2017})}\BibitemShut {NoStop}%
\bibitem [{\citenamefont {Chow}\ \emph {et~al.}(2011)\citenamefont {Chow},
  \citenamefont {C{\'o}rcoles}, \citenamefont {Gambetta}, \citenamefont
  {Rigetti}, \citenamefont {Johnson}, \citenamefont {Smolin}, \citenamefont
  {Rozen}, \citenamefont {Keefe}, \citenamefont {Rothwell}, \citenamefont
  {Ketchen},\ and\ \citenamefont {Steffen}}]{chow_CR_2011}%
  \BibitemOpen
  \bibfield  {author} {\bibinfo {author} {\bibfnamefont {J.~M.}\ \bibnamefont
  {Chow}}, \bibinfo {author} {\bibfnamefont {A.~D.}\ \bibnamefont
  {C{\'o}rcoles}}, \bibinfo {author} {\bibfnamefont {J.~M.}\ \bibnamefont
  {Gambetta}}, \bibinfo {author} {\bibfnamefont {C.}~\bibnamefont {Rigetti}},
  \bibinfo {author} {\bibfnamefont {B.~R.}\ \bibnamefont {Johnson}}, \bibinfo
  {author} {\bibfnamefont {J.~A.}\ \bibnamefont {Smolin}}, \bibinfo {author}
  {\bibfnamefont {J.~R.}\ \bibnamefont {Rozen}}, \bibinfo {author}
  {\bibfnamefont {G.~A.}\ \bibnamefont {Keefe}}, \bibinfo {author}
  {\bibfnamefont {M.~B.}\ \bibnamefont {Rothwell}}, \bibinfo {author}
  {\bibfnamefont {M.~B.}\ \bibnamefont {Ketchen}}, \ and\ \bibinfo {author}
  {\bibfnamefont {M.}~\bibnamefont {Steffen}},\ }\href {\doibase
  10.1103/PhysRevLett.107.080502} {\bibfield  {journal} {\bibinfo  {journal}
  {Physical Review Letters}\ }\textbf {\bibinfo {volume} {107}},\ \bibinfo
  {pages} {080502} (\bibinfo {year} {2011})}\BibitemShut {NoStop}%
\bibitem [{\citenamefont {Rigetti}\ and\ \citenamefont
  {Devoret}(2010)}]{rigetti}%
  \BibitemOpen
  \bibfield  {author} {\bibinfo {author} {\bibfnamefont {C.}~\bibnamefont
  {Rigetti}}\ and\ \bibinfo {author} {\bibfnamefont {M.}~\bibnamefont
  {Devoret}},\ }\href {\doibase 10.1103/PhysRevB.81.134507} {\bibfield
  {journal} {\bibinfo  {journal} {Phys. Rev. B}\ }\textbf {\bibinfo {volume}
  {81}},\ \bibinfo {pages} {134507} (\bibinfo {year} {2010})}\BibitemShut
  {NoStop}%
\bibitem [{\citenamefont {Sheldon}\ \emph
  {et~al.}(2016{\natexlab{b}})\citenamefont {Sheldon}, \citenamefont {Magesan},
  \citenamefont {Chow},\ and\ \citenamefont {Gambetta}}]{sheldonCR}%
  \BibitemOpen
  \bibfield  {author} {\bibinfo {author} {\bibfnamefont {S.}~\bibnamefont
  {Sheldon}}, \bibinfo {author} {\bibfnamefont {E.}~\bibnamefont {Magesan}},
  \bibinfo {author} {\bibfnamefont {J.~M.}\ \bibnamefont {Chow}}, \ and\
  \bibinfo {author} {\bibfnamefont {J.~M.}\ \bibnamefont {Gambetta}},\ }\href
  {\doibase 10.1103/PhysRevA.93.060302} {\bibfield  {journal} {\bibinfo
  {journal} {Phys. Rev. A}\ }\textbf {\bibinfo {volume} {93}},\ \bibinfo
  {pages} {060302} (\bibinfo {year} {2016}{\natexlab{b}})}\BibitemShut
  {NoStop}%
\bibitem [{qvi()}]{qvibm}%
  \BibitemOpen
  \href@noop {} {\enquote {\bibinfo {title} {Cramming more power into a quantum
  device},}\ }\bibinfo {howpublished}
  {\url{https://www.ibm.com/blogs/research/2019/03/power-quantum-device/}},\
  \bibinfo {note} {accessed: 2019-03-04}\BibitemShut {NoStop}%
\bibitem [{\citenamefont {Baum}\ \emph {et~al.}(1985)\citenamefont {Baum},
  \citenamefont {Munowitz}, \citenamefont {Garroway},\ and\ \citenamefont
  {Pines}}]{baum85jcp}%
  \BibitemOpen
  \bibfield  {author} {\bibinfo {author} {\bibfnamefont {J.}~\bibnamefont
  {Baum}}, \bibinfo {author} {\bibfnamefont {M.}~\bibnamefont {Munowitz}},
  \bibinfo {author} {\bibfnamefont {A.~N.}\ \bibnamefont {Garroway}}, \ and\
  \bibinfo {author} {\bibfnamefont {A.}~\bibnamefont {Pines}},\ }\href
  {\doibase 10.1063/1.449344} {\bibfield  {journal} {\bibinfo  {journal} {The
  Journal of Chemical Physics}\ }\textbf {\bibinfo {volume} {83}},\ \bibinfo
  {pages} {2015} (\bibinfo {year} {1985})},\ \Eprint
  {http://arxiv.org/abs/https://doi.org/10.1063/1.449344}
  {https://doi.org/10.1063/1.449344} \BibitemShut {NoStop}%
\bibitem [{\citenamefont {Wei}\ \emph {et~al.}(2018)\citenamefont {Wei},
  \citenamefont {Ramanathan},\ and\ \citenamefont {Cappellaro}}]{wei18prl}%
  \BibitemOpen
  \bibfield  {author} {\bibinfo {author} {\bibfnamefont {K.~X.}\ \bibnamefont
  {Wei}}, \bibinfo {author} {\bibfnamefont {C.}~\bibnamefont {Ramanathan}}, \
  and\ \bibinfo {author} {\bibfnamefont {P.}~\bibnamefont {Cappellaro}},\
  }\href {\doibase 10.1103/PhysRevLett.120.070501} {\bibfield  {journal}
  {\bibinfo  {journal} {Phys. Rev. Lett.}\ }\textbf {\bibinfo {volume} {120}},\
  \bibinfo {pages} {070501} (\bibinfo {year} {2018})}\BibitemShut {NoStop}%
\bibitem [{\citenamefont {G{\"a}rttner}\ \emph {et~al.}(2017)\citenamefont
  {G{\"a}rttner}, \citenamefont {Bohnet}, \citenamefont {Safavi-Naini},
  \citenamefont {Wall}, \citenamefont {Bollinger},\ and\ \citenamefont
  {Rey}}]{garttner17nphy}%
  \BibitemOpen
  \bibfield  {author} {\bibinfo {author} {\bibfnamefont {M.}~\bibnamefont
  {G{\"a}rttner}}, \bibinfo {author} {\bibfnamefont {J.~G.}\ \bibnamefont
  {Bohnet}}, \bibinfo {author} {\bibfnamefont {A.}~\bibnamefont
  {Safavi-Naini}}, \bibinfo {author} {\bibfnamefont {M.~L.}\ \bibnamefont
  {Wall}}, \bibinfo {author} {\bibfnamefont {J.~J.}\ \bibnamefont {Bollinger}},
  \ and\ \bibinfo {author} {\bibfnamefont {A.~M.}\ \bibnamefont {Rey}},\ }\href
  {https://doi.org/10.1038/nphys4119} {\bibfield  {journal} {\bibinfo
  {journal} {Nature Physics}\ }\textbf {\bibinfo {volume} {13}},\ \bibinfo
  {pages} {781 EP } (\bibinfo {year} {2017})},\ \bibinfo {note}
  {article}\BibitemShut {NoStop}%
\bibitem [{\citenamefont {Cappellaro}\ \emph {et~al.}(2005)\citenamefont
  {Cappellaro}, \citenamefont {Emerson}, \citenamefont {Boulant}, \citenamefont
  {Ramanathan}, \citenamefont {Lloyd},\ and\ \citenamefont
  {Cory}}]{cappellaro05prl}%
  \BibitemOpen
  \bibfield  {author} {\bibinfo {author} {\bibfnamefont {P.}~\bibnamefont
  {Cappellaro}}, \bibinfo {author} {\bibfnamefont {J.}~\bibnamefont {Emerson}},
  \bibinfo {author} {\bibfnamefont {N.}~\bibnamefont {Boulant}}, \bibinfo
  {author} {\bibfnamefont {C.}~\bibnamefont {Ramanathan}}, \bibinfo {author}
  {\bibfnamefont {S.}~\bibnamefont {Lloyd}}, \ and\ \bibinfo {author}
  {\bibfnamefont {D.~G.}\ \bibnamefont {Cory}},\ }\href {\doibase
  10.1103/PhysRevLett.94.020502} {\bibfield  {journal} {\bibinfo  {journal}
  {Phys. Rev. Lett.}\ }\textbf {\bibinfo {volume} {94}},\ \bibinfo {pages}
  {020502} (\bibinfo {year} {2005})}\BibitemShut {NoStop}%
\bibitem [{\citenamefont {Bollinger}\ \emph {et~al.}(1996)\citenamefont
  {Bollinger}, \citenamefont {Itano}, \citenamefont {Wineland},\ and\
  \citenamefont {Heinzen}}]{bollinger96a}%
  \BibitemOpen
  \bibfield  {author} {\bibinfo {author} {\bibfnamefont {J.~J.}\ \bibnamefont
  {Bollinger}}, \bibinfo {author} {\bibfnamefont {W.~M.}\ \bibnamefont
  {Itano}}, \bibinfo {author} {\bibfnamefont {D.~J.}\ \bibnamefont {Wineland}},
  \ and\ \bibinfo {author} {\bibfnamefont {D.~J.}\ \bibnamefont {Heinzen}},\
  }\href {\doibase 10.1103/PhysRevA.54.R4649} {\bibfield  {journal} {\bibinfo
  {journal} {Phys. Rev. A}\ }\textbf {\bibinfo {volume} {54}},\ \bibinfo
  {pages} {R4649} (\bibinfo {year} {1996})}\BibitemShut {NoStop}%
\bibitem [{\citenamefont {Leibfried}\ \emph {et~al.}(2004)\citenamefont
  {Leibfried}, \citenamefont {Barrett}, \citenamefont {Schaetz}, \citenamefont
  {Britton}, \citenamefont {Chiaverini}, \citenamefont {Itano}, \citenamefont
  {Jost}, \citenamefont {Langer},\ and\ \citenamefont
  {Wineland}}]{Leibfried1476}%
  \BibitemOpen
  \bibfield  {author} {\bibinfo {author} {\bibfnamefont {D.}~\bibnamefont
  {Leibfried}}, \bibinfo {author} {\bibfnamefont {M.~D.}\ \bibnamefont
  {Barrett}}, \bibinfo {author} {\bibfnamefont {T.}~\bibnamefont {Schaetz}},
  \bibinfo {author} {\bibfnamefont {J.}~\bibnamefont {Britton}}, \bibinfo
  {author} {\bibfnamefont {J.}~\bibnamefont {Chiaverini}}, \bibinfo {author}
  {\bibfnamefont {W.~M.}\ \bibnamefont {Itano}}, \bibinfo {author}
  {\bibfnamefont {J.~D.}\ \bibnamefont {Jost}}, \bibinfo {author}
  {\bibfnamefont {C.}~\bibnamefont {Langer}}, \ and\ \bibinfo {author}
  {\bibfnamefont {D.~J.}\ \bibnamefont {Wineland}},\ }\href {\doibase
  10.1126/science.1097576} {\bibfield  {journal} {\bibinfo  {journal}
  {Science}\ }\textbf {\bibinfo {volume} {304}},\ \bibinfo {pages} {1476}
  (\bibinfo {year} {2004})}\BibitemShut {NoStop}%
\bibitem [{\citenamefont {Giovannetti}\ \emph {et~al.}(2004)\citenamefont
  {Giovannetti}, \citenamefont {Lloyd},\ and\ \citenamefont
  {Maccone}}]{Giovannetti1330}%
  \BibitemOpen
  \bibfield  {author} {\bibinfo {author} {\bibfnamefont {V.}~\bibnamefont
  {Giovannetti}}, \bibinfo {author} {\bibfnamefont {S.}~\bibnamefont {Lloyd}},
  \ and\ \bibinfo {author} {\bibfnamefont {L.}~\bibnamefont {Maccone}},\ }\href
  {\doibase 10.1126/science.1104149} {\bibfield  {journal} {\bibinfo  {journal}
  {Science}\ }\textbf {\bibinfo {volume} {306}},\ \bibinfo {pages} {1330}
  (\bibinfo {year} {2004})}\BibitemShut {NoStop}%
\bibitem [{\citenamefont {Gühne}\ and\ \citenamefont
  {Tóth}(2009)}]{guhne09pr}%
  \BibitemOpen
  \bibfield  {author} {\bibinfo {author} {\bibfnamefont {O.}~\bibnamefont
  {Gühne}}\ and\ \bibinfo {author} {\bibfnamefont {G.}~\bibnamefont {Tóth}},\
  }\href {\doibase https://doi.org/10.1016/j.physrep.2009.02.004} {\bibfield
  {journal} {\bibinfo  {journal} {Physics Reports}\ }\textbf {\bibinfo {volume}
  {474}},\ \bibinfo {pages} {1 } (\bibinfo {year} {2009})}\BibitemShut
  {NoStop}%
\bibitem [{\citenamefont {Gühne}\ and\ \citenamefont
  {Seevinck}(2010)}]{Guhne_njp}%
  \BibitemOpen
  \bibfield  {author} {\bibinfo {author} {\bibfnamefont {O.}~\bibnamefont
  {Gühne}}\ and\ \bibinfo {author} {\bibfnamefont {M.}~\bibnamefont
  {Seevinck}},\ }\href {\doibase 10.1088/1367-2630/12/5/053002} {\bibfield
  {journal} {\bibinfo  {journal} {New Journal of Physics}\ }\textbf {\bibinfo
  {volume} {12}},\ \bibinfo {pages} {053002} (\bibinfo {year}
  {2010})}\BibitemShut {NoStop}%
\bibitem [{\citenamefont {McKay}\ \emph
  {et~al.}(2017{\natexlab{b}})\citenamefont {McKay}, \citenamefont {Wood},
  \citenamefont {Sheldon}, \citenamefont {Chow},\ and\ \citenamefont
  {Gambetta}}]{McKay2016b}%
  \BibitemOpen
  \bibfield  {author} {\bibinfo {author} {\bibfnamefont {D.~C.}\ \bibnamefont
  {McKay}}, \bibinfo {author} {\bibfnamefont {C.~J.}\ \bibnamefont {Wood}},
  \bibinfo {author} {\bibfnamefont {S.}~\bibnamefont {Sheldon}}, \bibinfo
  {author} {\bibfnamefont {J.~M.}\ \bibnamefont {Chow}}, \ and\ \bibinfo
  {author} {\bibfnamefont {J.~M.}\ \bibnamefont {Gambetta}},\ }\href {\doibase
  10.1103/PhysRevA.96.022330} {\bibfield  {journal} {\bibinfo  {journal} {Phys.
  Rev. A}\ }\textbf {\bibinfo {volume} {96}},\ \bibinfo {pages} {022330}
  (\bibinfo {year} {2017}{\natexlab{b}})}\BibitemShut {NoStop}%
\bibitem [{\citenamefont {Sackett}\ \emph {et~al.}(2000)\citenamefont
  {Sackett}, \citenamefont {Kielpinski}, \citenamefont {King}, \citenamefont
  {Langer}, \citenamefont {Meyer}, \citenamefont {Myatt}, \citenamefont {Rowe},
  \citenamefont {Turchette}, \citenamefont {Itano}, \citenamefont {Wineland},\
  and\ \citenamefont {Monroe}}]{sackett00n}%
  \BibitemOpen
  \bibfield  {author} {\bibinfo {author} {\bibfnamefont {C.~A.}\ \bibnamefont
  {Sackett}}, \bibinfo {author} {\bibfnamefont {D.}~\bibnamefont {Kielpinski}},
  \bibinfo {author} {\bibfnamefont {B.~E.}\ \bibnamefont {King}}, \bibinfo
  {author} {\bibfnamefont {C.}~\bibnamefont {Langer}}, \bibinfo {author}
  {\bibfnamefont {V.}~\bibnamefont {Meyer}}, \bibinfo {author} {\bibfnamefont
  {C.~J.}\ \bibnamefont {Myatt}}, \bibinfo {author} {\bibfnamefont
  {M.}~\bibnamefont {Rowe}}, \bibinfo {author} {\bibfnamefont {Q.~A.}\
  \bibnamefont {Turchette}}, \bibinfo {author} {\bibfnamefont {W.~M.}\
  \bibnamefont {Itano}}, \bibinfo {author} {\bibfnamefont {D.~J.}\ \bibnamefont
  {Wineland}}, \ and\ \bibinfo {author} {\bibfnamefont {C.}~\bibnamefont
  {Monroe}},\ }\href {https://doi.org/10.1038/35005011} {\bibfield  {journal}
  {\bibinfo  {journal} {Nature}\ }\textbf {\bibinfo {volume} {404}},\ \bibinfo
  {pages} {256 EP } (\bibinfo {year} {2000})}\BibitemShut {NoStop}%
\bibitem [{\citenamefont {Leibfried}\ \emph {et~al.}(2005)\citenamefont
  {Leibfried}, \citenamefont {Knill}, \citenamefont {Seidelin}, \citenamefont
  {Britton}, \citenamefont {Blakestad}, \citenamefont {Chiaverini},
  \citenamefont {Hume}, \citenamefont {Itano}, \citenamefont {Jost},
  \citenamefont {Langer}, \citenamefont {Ozeri}, \citenamefont {Reichle},\ and\
  \citenamefont {Wineland}}]{leibfried05n}%
  \BibitemOpen
  \bibfield  {author} {\bibinfo {author} {\bibfnamefont {D.}~\bibnamefont
  {Leibfried}}, \bibinfo {author} {\bibfnamefont {E.}~\bibnamefont {Knill}},
  \bibinfo {author} {\bibfnamefont {S.}~\bibnamefont {Seidelin}}, \bibinfo
  {author} {\bibfnamefont {J.}~\bibnamefont {Britton}}, \bibinfo {author}
  {\bibfnamefont {R.~B.}\ \bibnamefont {Blakestad}}, \bibinfo {author}
  {\bibfnamefont {J.}~\bibnamefont {Chiaverini}}, \bibinfo {author}
  {\bibfnamefont {D.~B.}\ \bibnamefont {Hume}}, \bibinfo {author}
  {\bibfnamefont {W.~M.}\ \bibnamefont {Itano}}, \bibinfo {author}
  {\bibfnamefont {J.~D.}\ \bibnamefont {Jost}}, \bibinfo {author}
  {\bibfnamefont {C.}~\bibnamefont {Langer}}, \bibinfo {author} {\bibfnamefont
  {R.}~\bibnamefont {Ozeri}}, \bibinfo {author} {\bibfnamefont
  {R.}~\bibnamefont {Reichle}}, \ and\ \bibinfo {author} {\bibfnamefont
  {D.~J.}\ \bibnamefont {Wineland}},\ }\href
  {https://doi.org/10.1038/nature04251} {\bibfield  {journal} {\bibinfo
  {journal} {Nature}\ }\textbf {\bibinfo {volume} {438}},\ \bibinfo {pages}
  {639 EP } (\bibinfo {year} {2005})}\BibitemShut {NoStop}%
\bibitem [{\citenamefont {Hahn}(1950)}]{hahn50}%
  \BibitemOpen
  \bibfield  {author} {\bibinfo {author} {\bibfnamefont {E.~L.}\ \bibnamefont
  {Hahn}},\ }\href {\doibase 10.1103/PhysRev.80.580} {\bibfield  {journal}
  {\bibinfo  {journal} {Phys. Rev.}\ }\textbf {\bibinfo {volume} {80}},\
  \bibinfo {pages} {580} (\bibinfo {year} {1950})}\BibitemShut {NoStop}%
\bibitem [{\citenamefont {{Cruz}}\ \emph {et~al.}(2018)\citenamefont {{Cruz}},
  \citenamefont {{Fournier}}, \citenamefont {{Gremion}}, \citenamefont
  {{Jeannerot}}, \citenamefont {{Komagata}}, \citenamefont {{Tosic}},
  \citenamefont {{Thiesbrummel}}, \citenamefont {{Chan}}, \citenamefont
  {{Macris}}, \citenamefont {{Dupertuis}},\ and\ \citenamefont
  {{Javerzac-Galy}}}]{cruz18x}%
  \BibitemOpen
  \bibfield  {author} {\bibinfo {author} {\bibfnamefont {D.}~\bibnamefont
  {{Cruz}}}, \bibinfo {author} {\bibfnamefont {R.}~\bibnamefont {{Fournier}}},
  \bibinfo {author} {\bibfnamefont {F.}~\bibnamefont {{Gremion}}}, \bibinfo
  {author} {\bibfnamefont {A.}~\bibnamefont {{Jeannerot}}}, \bibinfo {author}
  {\bibfnamefont {K.}~\bibnamefont {{Komagata}}}, \bibinfo {author}
  {\bibfnamefont {T.}~\bibnamefont {{Tosic}}}, \bibinfo {author} {\bibfnamefont
  {J.}~\bibnamefont {{Thiesbrummel}}}, \bibinfo {author} {\bibfnamefont
  {C.~L.}\ \bibnamefont {{Chan}}}, \bibinfo {author} {\bibfnamefont
  {N.}~\bibnamefont {{Macris}}}, \bibinfo {author} {\bibfnamefont {M.-A.}\
  \bibnamefont {{Dupertuis}}}, \ and\ \bibinfo {author} {\bibfnamefont
  {C.}~\bibnamefont {{Javerzac-Galy}}},\ }\href@noop {} {\bibfield  {journal}
  {\bibinfo  {journal} {arXiv e-prints}\ } (\bibinfo {year} {2018})},\ \Eprint
  {http://arxiv.org/abs/1807.05572} {arXiv:1807.05572 [quant-ph]} \BibitemShut
  {NoStop}%
\bibitem [{\citenamefont {Anderson}\ \emph {et~al.}(2019)\citenamefont
  {Anderson}, \citenamefont {Dahl},\ and\ \citenamefont
  {Vandenberghe}}]{cvxopt}%
  \BibitemOpen
  \bibfield  {author} {\bibinfo {author} {\bibfnamefont {M.~S.}\ \bibnamefont
  {Anderson}}, \bibinfo {author} {\bibfnamefont {J.}~\bibnamefont {Dahl}}, \
  and\ \bibinfo {author} {\bibfnamefont {L.}~\bibnamefont {Vandenberghe}},\
  }\href@noop {} {\enquote {\bibinfo {title} {{CVXOPT}: A python package for
  convex optimization, version 1.2.3},}\ }\bibinfo {howpublished}
  {\url{https://cvxopt.org/}} (\bibinfo {year} {2019})\BibitemShut {NoStop}%
\bibitem [{\citenamefont {Smolin}\ \emph {et~al.}(2012)\citenamefont {Smolin},
  \citenamefont {Gambetta},\ and\ \citenamefont {Smith}}]{Smolin12}%
  \BibitemOpen
  \bibfield  {author} {\bibinfo {author} {\bibfnamefont {J.~A.}\ \bibnamefont
  {Smolin}}, \bibinfo {author} {\bibfnamefont {J.~M.}\ \bibnamefont
  {Gambetta}}, \ and\ \bibinfo {author} {\bibfnamefont {G.}~\bibnamefont
  {Smith}},\ }\href {\doibase 10.1103/PhysRevLett.108.070502} {\bibfield
  {journal} {\bibinfo  {journal} {Phys. Rev. Lett.}\ }\textbf {\bibinfo
  {volume} {108}},\ \bibinfo {pages} {070502} (\bibinfo {year}
  {2012})}\BibitemShut {NoStop}%
\bibitem [{\citenamefont {Hein}\ \emph {et~al.}(2004)\citenamefont {Hein},
  \citenamefont {Eisert},\ and\ \citenamefont {Briegel}}]{hein04a}%
  \BibitemOpen
  \bibfield  {author} {\bibinfo {author} {\bibfnamefont {M.}~\bibnamefont
  {Hein}}, \bibinfo {author} {\bibfnamefont {J.}~\bibnamefont {Eisert}}, \ and\
  \bibinfo {author} {\bibfnamefont {H.~J.}\ \bibnamefont {Briegel}},\ }\href
  {\doibase 10.1103/PhysRevA.69.062311} {\bibfield  {journal} {\bibinfo
  {journal} {Phys. Rev. A}\ }\textbf {\bibinfo {volume} {69}},\ \bibinfo
  {pages} {062311} (\bibinfo {year} {2004})}\BibitemShut {NoStop}%
\bibitem [{\citenamefont {Aleksandrowicz}\ \emph {et~al.}(2019)\citenamefont
  {Aleksandrowicz}, \citenamefont {Alexander}, \citenamefont {Barkoutsos},
  \citenamefont {Bello}, \citenamefont {Ben-Haim}, \citenamefont {Bucher},
  \citenamefont {Cabrera-Hern{\'a}dez}, \citenamefont {Carballo-Franquis},
  \citenamefont {Chen}, \citenamefont {Chen}, \citenamefont {Chow},
  \citenamefont {C{\'o}rcoles-Gonzales}, \citenamefont {Cross}, \citenamefont
  {Cross}, \citenamefont {Cruz-Benito}, \citenamefont {Culver}, \citenamefont
  {Gonz{\'a}lez}, \citenamefont {Torre}, \citenamefont {Ding}, \citenamefont
  {Dumitrescu}, \citenamefont {Duran}, \citenamefont {Eendebak}, \citenamefont
  {Everitt}, \citenamefont {Sertage}, \citenamefont {Frisch}, \citenamefont
  {Fuhrer}, \citenamefont {Gambetta}, \citenamefont {Gago}, \citenamefont
  {Gomez-Mosquera}, \citenamefont {Greenberg}, \citenamefont {Hamamura},
  \citenamefont {Havlicek}, \citenamefont {Hellmers}, \citenamefont {Herok},
  \citenamefont {Horii}, \citenamefont {Hu}, \citenamefont {Imamichi},
  \citenamefont {Itoko}, \citenamefont {Javadi-Abhari}, \citenamefont
  {Kanazawa}, \citenamefont {Karazeev}, \citenamefont {Krsulich}, \citenamefont
  {Liu}, \citenamefont {Luh}, \citenamefont {Maeng}, \citenamefont {Marques},
  \citenamefont {Mart{\'\i}n-Fern{\'a}ndez}, \citenamefont {McClure},
  \citenamefont {McKay}, \citenamefont {Meesala}, \citenamefont {Mezzacapo},
  \citenamefont {Moll}, \citenamefont {Rodr{\'\i}guez}, \citenamefont
  {Nannicini}, \citenamefont {Nation}, \citenamefont {Ollitrault},
  \citenamefont {O'Riordan}, \citenamefont {Paik}, \citenamefont {P{\'e}rez},
  \citenamefont {Phan}, \citenamefont {Pistoia}, \citenamefont {Prutyanov},
  \citenamefont {Reuter}, \citenamefont {Rice}, \citenamefont {Davila},
  \citenamefont {Rudy}, \citenamefont {Ryu}, \citenamefont {Sathaye},
  \citenamefont {Schnabel}, \citenamefont {Schoute}, \citenamefont {Setia},
  \citenamefont {Shi}, \citenamefont {Silva}, \citenamefont {Siraichi},
  \citenamefont {Sivarajah}, \citenamefont {Smolin}, \citenamefont {Soeken},
  \citenamefont {Takahashi}, \citenamefont {Tavernelli}, \citenamefont
  {Taylor}, \citenamefont {Taylour}, \citenamefont {Trabing}, \citenamefont
  {Treinish}, \citenamefont {Turner}, \citenamefont {Vogt-Lee}, \citenamefont
  {Vuillot}, \citenamefont {Wildstrom}, \citenamefont {Wilson}, \citenamefont
  {Winston}, \citenamefont {Wood}, \citenamefont {Wood}, \citenamefont
  {W{\"o}rner}, \citenamefont {Akhalwaya},\ and\ \citenamefont
  {Zoufal}}]{Qiskit19}%
  \BibitemOpen
  \bibfield  {author} {\bibinfo {author} {\bibfnamefont {G.}~\bibnamefont
  {Aleksandrowicz}}, \bibinfo {author} {\bibfnamefont {T.}~\bibnamefont
  {Alexander}}, \bibinfo {author} {\bibfnamefont {P.}~\bibnamefont
  {Barkoutsos}}, \bibinfo {author} {\bibfnamefont {L.}~\bibnamefont {Bello}},
  \bibinfo {author} {\bibfnamefont {Y.}~\bibnamefont {Ben-Haim}}, \bibinfo
  {author} {\bibfnamefont {D.}~\bibnamefont {Bucher}}, \bibinfo {author}
  {\bibfnamefont {F.~J.}\ \bibnamefont {Cabrera-Hern{\'a}dez}}, \bibinfo
  {author} {\bibfnamefont {J.}~\bibnamefont {Carballo-Franquis}}, \bibinfo
  {author} {\bibfnamefont {A.}~\bibnamefont {Chen}}, \bibinfo {author}
  {\bibfnamefont {C.-F.}\ \bibnamefont {Chen}}, \bibinfo {author}
  {\bibfnamefont {J.~M.}\ \bibnamefont {Chow}}, \bibinfo {author}
  {\bibfnamefont {A.~D.}\ \bibnamefont {C{\'o}rcoles-Gonzales}}, \bibinfo
  {author} {\bibfnamefont {A.~J.}\ \bibnamefont {Cross}}, \bibinfo {author}
  {\bibfnamefont {A.}~\bibnamefont {Cross}}, \bibinfo {author} {\bibfnamefont
  {J.}~\bibnamefont {Cruz-Benito}}, \bibinfo {author} {\bibfnamefont
  {C.}~\bibnamefont {Culver}}, \bibinfo {author} {\bibfnamefont {S.~D. L.~P.}\
  \bibnamefont {Gonz{\'a}lez}}, \bibinfo {author} {\bibfnamefont {E.~D.~L.}\
  \bibnamefont {Torre}}, \bibinfo {author} {\bibfnamefont {D.}~\bibnamefont
  {Ding}}, \bibinfo {author} {\bibfnamefont {E.}~\bibnamefont {Dumitrescu}},
  \bibinfo {author} {\bibfnamefont {I.}~\bibnamefont {Duran}}, \bibinfo
  {author} {\bibfnamefont {P.}~\bibnamefont {Eendebak}}, \bibinfo {author}
  {\bibfnamefont {M.}~\bibnamefont {Everitt}}, \bibinfo {author} {\bibfnamefont
  {I.~F.}\ \bibnamefont {Sertage}}, \bibinfo {author} {\bibfnamefont
  {A.}~\bibnamefont {Frisch}}, \bibinfo {author} {\bibfnamefont
  {A.}~\bibnamefont {Fuhrer}}, \bibinfo {author} {\bibfnamefont
  {J.}~\bibnamefont {Gambetta}}, \bibinfo {author} {\bibfnamefont {B.~G.}\
  \bibnamefont {Gago}}, \bibinfo {author} {\bibfnamefont {J.}~\bibnamefont
  {Gomez-Mosquera}}, \bibinfo {author} {\bibfnamefont {D.}~\bibnamefont
  {Greenberg}}, \bibinfo {author} {\bibfnamefont {I.}~\bibnamefont {Hamamura}},
  \bibinfo {author} {\bibfnamefont {V.}~\bibnamefont {Havlicek}}, \bibinfo
  {author} {\bibfnamefont {J.}~\bibnamefont {Hellmers}}, \bibinfo {author}
  {\bibfnamefont {{\L}.}~\bibnamefont {Herok}}, \bibinfo {author}
  {\bibfnamefont {H.}~\bibnamefont {Horii}}, \bibinfo {author} {\bibfnamefont
  {S.}~\bibnamefont {Hu}}, \bibinfo {author} {\bibfnamefont {T.}~\bibnamefont
  {Imamichi}}, \bibinfo {author} {\bibfnamefont {T.}~\bibnamefont {Itoko}},
  \bibinfo {author} {\bibfnamefont {A.}~\bibnamefont {Javadi-Abhari}}, \bibinfo
  {author} {\bibfnamefont {N.}~\bibnamefont {Kanazawa}}, \bibinfo {author}
  {\bibfnamefont {A.}~\bibnamefont {Karazeev}}, \bibinfo {author}
  {\bibfnamefont {K.}~\bibnamefont {Krsulich}}, \bibinfo {author}
  {\bibfnamefont {P.}~\bibnamefont {Liu}}, \bibinfo {author} {\bibfnamefont
  {Y.}~\bibnamefont {Luh}}, \bibinfo {author} {\bibfnamefont {Y.}~\bibnamefont
  {Maeng}}, \bibinfo {author} {\bibfnamefont {M.}~\bibnamefont {Marques}},
  \bibinfo {author} {\bibfnamefont {F.~J.}\ \bibnamefont
  {Mart{\'\i}n-Fern{\'a}ndez}}, \bibinfo {author} {\bibfnamefont {D.~T.}\
  \bibnamefont {McClure}}, \bibinfo {author} {\bibfnamefont {D.}~\bibnamefont
  {McKay}}, \bibinfo {author} {\bibfnamefont {S.}~\bibnamefont {Meesala}},
  \bibinfo {author} {\bibfnamefont {A.}~\bibnamefont {Mezzacapo}}, \bibinfo
  {author} {\bibfnamefont {N.}~\bibnamefont {Moll}}, \bibinfo {author}
  {\bibfnamefont {D.~M.}\ \bibnamefont {Rodr{\'\i}guez}}, \bibinfo {author}
  {\bibfnamefont {G.}~\bibnamefont {Nannicini}}, \bibinfo {author}
  {\bibfnamefont {P.}~\bibnamefont {Nation}}, \bibinfo {author} {\bibfnamefont
  {P.}~\bibnamefont {Ollitrault}}, \bibinfo {author} {\bibfnamefont {L.~J.}\
  \bibnamefont {O'Riordan}}, \bibinfo {author} {\bibfnamefont {H.}~\bibnamefont
  {Paik}}, \bibinfo {author} {\bibfnamefont {J.}~\bibnamefont {P{\'e}rez}},
  \bibinfo {author} {\bibfnamefont {A.}~\bibnamefont {Phan}}, \bibinfo {author}
  {\bibfnamefont {M.}~\bibnamefont {Pistoia}}, \bibinfo {author} {\bibfnamefont
  {V.}~\bibnamefont {Prutyanov}}, \bibinfo {author} {\bibfnamefont
  {M.}~\bibnamefont {Reuter}}, \bibinfo {author} {\bibfnamefont
  {J.}~\bibnamefont {Rice}}, \bibinfo {author} {\bibfnamefont {A.~R.}\
  \bibnamefont {Davila}}, \bibinfo {author} {\bibfnamefont {R.~H.~P.}\
  \bibnamefont {Rudy}}, \bibinfo {author} {\bibfnamefont {M.}~\bibnamefont
  {Ryu}}, \bibinfo {author} {\bibfnamefont {N.}~\bibnamefont {Sathaye}},
  \bibinfo {author} {\bibfnamefont {C.}~\bibnamefont {Schnabel}}, \bibinfo
  {author} {\bibfnamefont {E.}~\bibnamefont {Schoute}}, \bibinfo {author}
  {\bibfnamefont {K.}~\bibnamefont {Setia}}, \bibinfo {author} {\bibfnamefont
  {Y.}~\bibnamefont {Shi}}, \bibinfo {author} {\bibfnamefont {A.}~\bibnamefont
  {Silva}}, \bibinfo {author} {\bibfnamefont {Y.}~\bibnamefont {Siraichi}},
  \bibinfo {author} {\bibfnamefont {S.}~\bibnamefont {Sivarajah}}, \bibinfo
  {author} {\bibfnamefont {J.~A.}\ \bibnamefont {Smolin}}, \bibinfo {author}
  {\bibfnamefont {M.}~\bibnamefont {Soeken}}, \bibinfo {author} {\bibfnamefont
  {H.}~\bibnamefont {Takahashi}}, \bibinfo {author} {\bibfnamefont
  {I.}~\bibnamefont {Tavernelli}}, \bibinfo {author} {\bibfnamefont
  {C.}~\bibnamefont {Taylor}}, \bibinfo {author} {\bibfnamefont
  {P.}~\bibnamefont {Taylour}}, \bibinfo {author} {\bibfnamefont
  {K.}~\bibnamefont {Trabing}}, \bibinfo {author} {\bibfnamefont
  {M.}~\bibnamefont {Treinish}}, \bibinfo {author} {\bibfnamefont
  {W.}~\bibnamefont {Turner}}, \bibinfo {author} {\bibfnamefont
  {D.}~\bibnamefont {Vogt-Lee}}, \bibinfo {author} {\bibfnamefont
  {C.}~\bibnamefont {Vuillot}}, \bibinfo {author} {\bibfnamefont {J.~A.}\
  \bibnamefont {Wildstrom}}, \bibinfo {author} {\bibfnamefont {J.}~\bibnamefont
  {Wilson}}, \bibinfo {author} {\bibfnamefont {E.}~\bibnamefont {Winston}},
  \bibinfo {author} {\bibfnamefont {C.}~\bibnamefont {Wood}}, \bibinfo {author}
  {\bibfnamefont {S.}~\bibnamefont {Wood}}, \bibinfo {author} {\bibfnamefont
  {S.}~\bibnamefont {W{\"o}rner}}, \bibinfo {author} {\bibfnamefont {I.~Y.}\
  \bibnamefont {Akhalwaya}}, \ and\ \bibinfo {author} {\bibfnamefont
  {C.}~\bibnamefont {Zoufal}},\ }\href {\doibase 10.5281/zenodo.2562110}
  {\enquote {\bibinfo {title} {Qiskit: An open-source framework for quantum
  computing},}\ } (\bibinfo {year} {2019})\BibitemShut {NoStop}%
\bibitem [{\citenamefont {Kaufmann}\ \emph {et~al.}(2017)\citenamefont
  {Kaufmann}, \citenamefont {Ruster}, \citenamefont {Schmiegelow},
  \citenamefont {Luda}, \citenamefont {Kaushal}, \citenamefont {Schulz},
  \citenamefont {von Lindenfels}, \citenamefont {Schmidt-Kaler},\ and\
  \citenamefont {Poschinger}}]{kaufmann17}%
  \BibitemOpen
  \bibfield  {author} {\bibinfo {author} {\bibfnamefont {H.}~\bibnamefont
  {Kaufmann}}, \bibinfo {author} {\bibfnamefont {T.}~\bibnamefont {Ruster}},
  \bibinfo {author} {\bibfnamefont {C.~T.}\ \bibnamefont {Schmiegelow}},
  \bibinfo {author} {\bibfnamefont {M.~A.}\ \bibnamefont {Luda}}, \bibinfo
  {author} {\bibfnamefont {V.}~\bibnamefont {Kaushal}}, \bibinfo {author}
  {\bibfnamefont {J.}~\bibnamefont {Schulz}}, \bibinfo {author} {\bibfnamefont
  {D.}~\bibnamefont {von Lindenfels}}, \bibinfo {author} {\bibfnamefont
  {F.}~\bibnamefont {Schmidt-Kaler}}, \ and\ \bibinfo {author} {\bibfnamefont
  {U.~G.}\ \bibnamefont {Poschinger}},\ }\href {\doibase
  10.1103/PhysRevLett.119.150503} {\bibfield  {journal} {\bibinfo  {journal}
  {Phys. Rev. Lett.}\ }\textbf {\bibinfo {volume} {119}},\ \bibinfo {pages}
  {150503} (\bibinfo {year} {2017})}\BibitemShut {NoStop}%
\bibitem [{\citenamefont {Takita}\ \emph {et~al.}(2017)\citenamefont {Takita},
  \citenamefont {Cross}, \citenamefont {C\'orcoles}, \citenamefont {Chow},\
  and\ \citenamefont {Gambetta}}]{Takita17}%
  \BibitemOpen
  \bibfield  {author} {\bibinfo {author} {\bibfnamefont {M.}~\bibnamefont
  {Takita}}, \bibinfo {author} {\bibfnamefont {A.~W.}\ \bibnamefont {Cross}},
  \bibinfo {author} {\bibfnamefont {A.~D.}\ \bibnamefont {C\'orcoles}},
  \bibinfo {author} {\bibfnamefont {J.~M.}\ \bibnamefont {Chow}}, \ and\
  \bibinfo {author} {\bibfnamefont {J.~M.}\ \bibnamefont {Gambetta}},\ }\href
  {\doibase 10.1103/PhysRevLett.119.180501} {\bibfield  {journal} {\bibinfo
  {journal} {Phys. Rev. Lett.}\ }\textbf {\bibinfo {volume} {119}},\ \bibinfo
  {pages} {180501} (\bibinfo {year} {2017})}\BibitemShut {NoStop}%
\bibitem [{\citenamefont {Ozaeta}\ and\ \citenamefont
  {McMahon}(2019)}]{Ozaeta_2019}%
  \BibitemOpen
  \bibfield  {author} {\bibinfo {author} {\bibfnamefont {A.}~\bibnamefont
  {Ozaeta}}\ and\ \bibinfo {author} {\bibfnamefont {P.~L.}\ \bibnamefont
  {McMahon}},\ }\href {\doibase 10.1088/2058-9565/ab13e5} {\bibfield  {journal}
  {\bibinfo  {journal} {Quantum Science and Technology}\ }\textbf {\bibinfo
  {volume} {4}},\ \bibinfo {pages} {025015} (\bibinfo {year}
  {2019})}\BibitemShut {NoStop}%
\bibitem [{\citenamefont {Temme}\ \emph {et~al.}(2017)\citenamefont {Temme},
  \citenamefont {Bravyi},\ and\ \citenamefont {Gambetta}}]{temme17l}%
  \BibitemOpen
  \bibfield  {author} {\bibinfo {author} {\bibfnamefont {K.}~\bibnamefont
  {Temme}}, \bibinfo {author} {\bibfnamefont {S.}~\bibnamefont {Bravyi}}, \
  and\ \bibinfo {author} {\bibfnamefont {J.~M.}\ \bibnamefont {Gambetta}},\
  }\href {\doibase 10.1103/PhysRevLett.119.180509} {\bibfield  {journal}
  {\bibinfo  {journal} {Phys. Rev. Lett.}\ }\textbf {\bibinfo {volume} {119}},\
  \bibinfo {pages} {180509} (\bibinfo {year} {2017})}\BibitemShut {NoStop}%
\bibitem [{\citenamefont {Kandala}\ \emph {et~al.}(2019)\citenamefont
  {Kandala}, \citenamefont {Temme}, \citenamefont {C{\'o}rcoles}, \citenamefont
  {Mezzacapo}, \citenamefont {Chow},\ and\ \citenamefont
  {Gambetta}}]{Kandala2019}%
  \BibitemOpen
  \bibfield  {author} {\bibinfo {author} {\bibfnamefont {A.}~\bibnamefont
  {Kandala}}, \bibinfo {author} {\bibfnamefont {K.}~\bibnamefont {Temme}},
  \bibinfo {author} {\bibfnamefont {A.~D.}\ \bibnamefont {C{\'o}rcoles}},
  \bibinfo {author} {\bibfnamefont {A.}~\bibnamefont {Mezzacapo}}, \bibinfo
  {author} {\bibfnamefont {J.~M.}\ \bibnamefont {Chow}}, \ and\ \bibinfo
  {author} {\bibfnamefont {J.~M.}\ \bibnamefont {Gambetta}},\ }\href {\doibase
  10.1038/s41586-019-1040-7} {\bibfield  {journal} {\bibinfo  {journal}
  {Nature}\ }\textbf {\bibinfo {volume} {567}},\ \bibinfo {pages} {491}
  (\bibinfo {year} {2019})}\BibitemShut {NoStop}%
\bibitem [{\citenamefont {Song}\ \emph {et~al.}(2019)\citenamefont {Song},
  \citenamefont {Xu}, \citenamefont {Li}, \citenamefont {Zhang}, \citenamefont
  {Zhang}, \citenamefont {Liu}, \citenamefont {Guo}, \citenamefont {Wang},
  \citenamefont {Ren}, \citenamefont {Hao}, \citenamefont {Feng}, \citenamefont
  {Fan}, \citenamefont {Zheng}, \citenamefont {Wang}, \citenamefont {Wang},\
  and\ \citenamefont {Zhu}}]{song19x}%
  \BibitemOpen
  \bibfield  {author} {\bibinfo {author} {\bibfnamefont {C.}~\bibnamefont
  {Song}}, \bibinfo {author} {\bibfnamefont {K.}~\bibnamefont {Xu}}, \bibinfo
  {author} {\bibfnamefont {H.}~\bibnamefont {Li}}, \bibinfo {author}
  {\bibfnamefont {Y.}~\bibnamefont {Zhang}}, \bibinfo {author} {\bibfnamefont
  {X.}~\bibnamefont {Zhang}}, \bibinfo {author} {\bibfnamefont
  {W.}~\bibnamefont {Liu}}, \bibinfo {author} {\bibfnamefont {Q.}~\bibnamefont
  {Guo}}, \bibinfo {author} {\bibfnamefont {Z.}~\bibnamefont {Wang}}, \bibinfo
  {author} {\bibfnamefont {W.}~\bibnamefont {Ren}}, \bibinfo {author}
  {\bibfnamefont {J.}~\bibnamefont {Hao}}, \bibinfo {author} {\bibfnamefont
  {H.}~\bibnamefont {Feng}}, \bibinfo {author} {\bibfnamefont {H.}~\bibnamefont
  {Fan}}, \bibinfo {author} {\bibfnamefont {D.}~\bibnamefont {Zheng}}, \bibinfo
  {author} {\bibfnamefont {D.}~\bibnamefont {Wang}}, \bibinfo {author}
  {\bibfnamefont {H.}~\bibnamefont {Wang}}, \ and\ \bibinfo {author}
  {\bibfnamefont {S.}~\bibnamefont {Zhu}},\ }\href@noop {} {\enquote {\bibinfo
  {title} {Observation of multi-component atomic schrödinger cat states of up
  to 20 qubits},}\ } (\bibinfo {year} {2019}),\ \Eprint
  {http://arxiv.org/abs/arXiv:1905.00320} {arXiv:1905.00320} \BibitemShut
  {NoStop}%
\bibitem [{\citenamefont {Omran}\ \emph {et~al.}(2019)\citenamefont {Omran},
  \citenamefont {Levine}, \citenamefont {Keesling}, \citenamefont {Semeghini},
  \citenamefont {Wang}, \citenamefont {Ebadi}, \citenamefont {Bernien},
  \citenamefont {Zibrov}, \citenamefont {Pichler}, \citenamefont {Choi},
  \citenamefont {Cui}, \citenamefont {Rossignolo}, \citenamefont {Rembold},
  \citenamefont {Montangero}, \citenamefont {Calarco}, \citenamefont {Endres},
  \citenamefont {Greiner}, \citenamefont {Vuletic},\ and\ \citenamefont
  {Lukin}}]{omran19}%
  \BibitemOpen
  \bibfield  {author} {\bibinfo {author} {\bibfnamefont {A.}~\bibnamefont
  {Omran}}, \bibinfo {author} {\bibfnamefont {H.}~\bibnamefont {Levine}},
  \bibinfo {author} {\bibfnamefont {A.}~\bibnamefont {Keesling}}, \bibinfo
  {author} {\bibfnamefont {G.}~\bibnamefont {Semeghini}}, \bibinfo {author}
  {\bibfnamefont {T.~T.}\ \bibnamefont {Wang}}, \bibinfo {author}
  {\bibfnamefont {S.}~\bibnamefont {Ebadi}}, \bibinfo {author} {\bibfnamefont
  {H.}~\bibnamefont {Bernien}}, \bibinfo {author} {\bibfnamefont {A.~S.}\
  \bibnamefont {Zibrov}}, \bibinfo {author} {\bibfnamefont {H.}~\bibnamefont
  {Pichler}}, \bibinfo {author} {\bibfnamefont {S.}~\bibnamefont {Choi}},
  \bibinfo {author} {\bibfnamefont {J.}~\bibnamefont {Cui}}, \bibinfo {author}
  {\bibfnamefont {M.}~\bibnamefont {Rossignolo}}, \bibinfo {author}
  {\bibfnamefont {P.}~\bibnamefont {Rembold}}, \bibinfo {author} {\bibfnamefont
  {S.}~\bibnamefont {Montangero}}, \bibinfo {author} {\bibfnamefont
  {T.}~\bibnamefont {Calarco}}, \bibinfo {author} {\bibfnamefont
  {M.}~\bibnamefont {Endres}}, \bibinfo {author} {\bibfnamefont
  {M.}~\bibnamefont {Greiner}}, \bibinfo {author} {\bibfnamefont
  {V.}~\bibnamefont {Vuletic}}, \ and\ \bibinfo {author} {\bibfnamefont
  {M.~D.}\ \bibnamefont {Lukin}},\ }\href@noop {} {\enquote {\bibinfo {title}
  {Generation and manipulation of schr\"{o}dinger cat states in rydberg atom
  arrays},}\ } (\bibinfo {year} {2019})\BibitemShut {NoStop}%
\bibitem [{\citenamefont {G\"arttner}\ \emph {et~al.}(2018)\citenamefont
  {G\"arttner}, \citenamefont {Hauke},\ and\ \citenamefont
  {Rey}}]{garttner18prl}%
  \BibitemOpen
  \bibfield  {author} {\bibinfo {author} {\bibfnamefont {M.}~\bibnamefont
  {G\"arttner}}, \bibinfo {author} {\bibfnamefont {P.}~\bibnamefont {Hauke}}, \
  and\ \bibinfo {author} {\bibfnamefont {A.~M.}\ \bibnamefont {Rey}},\ }\href
  {\doibase 10.1103/PhysRevLett.120.040402} {\bibfield  {journal} {\bibinfo
  {journal} {Phys. Rev. Lett.}\ }\textbf {\bibinfo {volume} {120}},\ \bibinfo
  {pages} {040402} (\bibinfo {year} {2018})}\BibitemShut {NoStop}%
\end{thebibliography}%
